\documentclass[aip,jmp,amsmath,amssymb,preprint]{revtex4-1}

\usepackage{graphicx}% Include figure files
\usepackage{dcolumn}% Align table columns on decimal point
\usepackage{bm}% bold math
\usepackage{hyperref}
\def\d{{\rm d}}
\def\ed{\d \!\!\!\! \d\, }
\def\ew{\wedge \!\!\!\!\!\! \wedge\,\, }
\def\ewe{\wedge \!\!\!\!\! \wedge\,\, }
\begin{document}

%\preprint{}

\title{Geometric Lagrangian approach to the physical degree of freedom count in field theory}

%\thanks{}

\author{Bogar D\'{\i}az}
\email{bdiaz@fcfm.buap.mx}
\affiliation{Facultad de Ciencias F\'isico Matem\'aticas, Benem\'erita Universidad Aut\'onoma de Puebla, Ciudad Universitaria, 72570, Puebla, Puebla, M\'exico}
\affiliation{Departamento de Matem\'aticas, Instituto de Ciencias, Benem\'erita Universidad Aut\'onoma de Puebla, Ciudad Universitaria, 72570, Puebla, Puebla, M\'exico}

\author{Merced Montesinos}
\email{merced@fis.cinvestav.mx}
\homepage[.URL:]{http://www.fis.cinvestav.mx/~merced}
\affiliation{Departamento de Matem\'aticas, Instituto de Ciencias, Benem\'erita Universidad Aut\'onoma de Puebla, Ciudad Universitaria, 72570, Puebla, Puebla, M\'exico}
\affiliation{Departamento de F\'isica, Cinvestav, Avenida Instituto Polit\'ecnico Nacional 2508, San Pedro Zacatenco, 07360, Gustavo A. Madero, Ciudad de M\'exico, M\'exico}

\date{\today}
\begin{abstract}
To circumvent some technical difficulties faced by the geometric Lagrangian approach to the physical degree of freedom count presented in the work of D\'iaz, Higuita, and Montesinos [\href{http://dx.doi.org/10.1063/1.4903183}{J. Math. Phys. {\bf 55}, 122901 (2014)}] that prevent its direct implementation to field theory, in this paper, we slightly modify the geometric Lagrangian approach in such a way that its resulting version works perfectly for field theory (and for particle systems, of course). As in previous work, the current approach also allows us to directly get the Lagrangian constraints, a new Lagrangian formula for the counting of the number of physical degrees of freedom, the gauge transformations, and the number of first- and second-class constraints for any action principle based on a Lagrangian depending on the fields and their first derivatives {\it without} performing any Dirac's canonical analysis. An advantage of this approach over the previous work is that it also allows us to handle the reducibility of the constraints and to get the off-shell gauge transformations. The theoretical framework is illustrated in 3-dimensional generalized general relativity (Palatini and Witten's exotic actions), Chern-Simons theory, $4$-dimensional BF theory, and $4$-dimensional general relativity given by Palatini's action with cosmological constant.
\end{abstract}

\maketitle

%%%%%%%%%%%%%%%%%%%%%%%%%%%%%%%%%%%
\section{Introduction}

Lagrangian and Hamiltonian methods are powerful tools to reveal the geometric structure behind the action principle for a given physical theory. Among the main theoretical approaches we find Dirac's canonical formalism \cite{Dirac1,Dirac2} and the  covariant canonical formalism (CCF) \cite{crnkovic, bombelli, Wald} from the physicists viewpoint, and the geometric formalism (GF) from the mathematicians viewpoint. \cite{Gotay, GotayNester, Hinds}

Dirac's canonical formalism allows us to get the classification of the Hamiltonian constraints,  the gauge transformations, the handling of the reducibility of the constraints, and the number of the physical degrees of freedom. In spite of its relevant properties, sometimes we physicists require to analyze a theory from the Lagrangian viewpoint without having to perform a detailed Dirac's canonical analysis. For that purpose, we might use the two Lagrangian approaches available in literature, CCF and GF, to get the wanted information. Nevertheless, none of these two Lagrangian approaches allows us to count the number of physical degrees of freedom, which is one of the goals when analyzing a physical theory. To fill out this gap, a Lagrangian method to count the number of physical degrees of freedom was presented in Ref.~\onlinecite{diaz}. Such a method was developed for particle systems and involves two approaches to make the counting: one geometric and the other non-geometric. 

The non-geometric approach of Ref.~\onlinecite{diaz} also works for field theory (in the same sense that the Hamiltonian analysis developed by Dirac works for both particle systems and field theory). Nonetheless, the geometric approach developed in Ref.~\onlinecite{diaz} faces some technical difficulties described in Section \ref{sec2} that prevent its direct implementation to field theory. To circumvent these difficulties, in this paper we report a modification of such a geometric Lagrangian approach in such a way that its resulting version works perfectly for both particle systems and field theory. Furthermore, its intrinsic geometric character makes this approach suitable to analyze geometric theories such as  generally covariant theories. 

This paper is organized as follows: Sec. \ref{sec2} contains the theoretical framework of the current approach, which is completely general and allows us to get the Lagrangian constraints, the number of physical degrees of freedom, the gauge transformations, and the number of first- and second-class constraints without performing any Dirac's canonical analysis. We have also included a detailed discussion about the role of the reducibility of the constraints and a procedure to obtain also the off-shell gauge transformations, which were not analyzed in Ref.~\onlinecite{diaz}. Section \ref{sec3} contains the implementation of the general theory to relevant physical field theories: 3-dimensional generalized general relativity (Palatini and Witten's exotic actions), Chern-Simons theory, 4-dimensional BF theory, and 4-dimensional general relativity written as Palatini's action with a cosmological constant. Section \ref{sec4} contains our concluding remarks. Also, for the sake of completeness and for the benefit of readers, in Appendix \ref{apen1} we have included a toy model that additionally illustrates the general theory of Sect. \ref{sec2} and allows us to display the differences of the current approach with that of Ref.~\onlinecite{diaz}. Finally, in Appendix \ref{apen2} we present the differences between the geometry considered along this paper and that of the CCF.  

%\\\\\\\\\\\\\\\\\\\\\\\Section II\\\\\\\\\\\\\\\\\\\\\\\\\\\\\\\\\\\\
\section{The geometric Lagrangian approach} \label{sec2}
We begin by recalling the map between the Lagrangian and Hamiltonian parameters reported in Ref.~\onlinecite{diaz}. If $N =$ number of configuration variables,  $l =$ number of Lagrangian constraints, $g =$ number of Noether identities (which coincides with the number of independent gauge parameters in the gauge transformations), and $e =$ number of gauge parameters plus its successive time derivatives, then we have that the number of first-class and second-class constraints in Dirac's canonical formalism is given by 
\begin{eqnarray}\label{map}
N_1 = e, \qquad N_2 = l+g-e,
\end{eqnarray} 
respectively. It is remarkable that we can count $N_1$ and $N_2$ using the Lagrangian formalism {\it only}. This knowledge leads to
\begin{eqnarray}\label{original}
&&\mbox{number of physical degrees of freedom} =N-\frac{1}{2}\left(l+g+e\right),
\end{eqnarray}
which is a Lagrangian formula to make the physical degree of freedom count (this formula was recently used to analyze certain higher derivative Lagrangian systems \cite{Klein, Klein2}). Furthermore, it was also shown in Ref.~\onlinecite{diaz} that the Lagrangian formula can be cast in geometric terms such as 
\begin{equation}
N-\frac{1}{2}\left(l+g+e\right) = 
\frac{1}{2} \operatorname{Rank} \left( \iota^*\Omega \right),\label{gl1}
\end{equation}
where $\iota$ denotes the inclusion over the Lagrangian constraint surface defined by all the Lagrangian constraints and $\Omega$ is the symplectic structure defined on the tangent space. Therefore, to make the counting of physical degrees of freedom, either we compute the right-hand side of (\ref{original}) (non-geometric approach) or we directly compute $\frac{1}{2} \operatorname{Rank} \left( \iota^*\Omega \right)$ (geometric approach). Both approaches lead to the same result. 

Regarding the non-geometric approach, the formula (\ref{original}) also holds for any classical field theory in the same sense that the formula for the physical degree of freedom count in Dirac's canonical formalism holds for both particle systems and field theory. Furthermore, this formula also works in the reducible case, in which case we must consider in (\ref{original}) only independent quantities. Along this work, we will explain how this can be done. On the other hand, for point particles the constraints depend on ($q, \dot{q}$) only, and so it is always possible to obtain $\iota^*\Omega$ and then calculate its rank in the geometric approach. But, in field theory the constraints can also depend on the spatial derivatives of the field variables. Due to this fact, it is impossible to calculate $\iota^*\Omega$  for field theory unless we can explicitly solve the Lagrangian constraints. Because of this, we provide a new way to calculate the right-hand side of (\ref{gl1}) without explicitly solving the constraints. This new formula is given by Eq. (\ref{pdfc}) of this paper. 

It is worth mentioning that following the geometric approach reported below, at the end, it is also possible to obtain $e$, $g$, and $l$, from which we could use (\ref{original}) for the counting if we wished. Even so, we want to mention that one advantage of calculating $\operatorname{Rank} \left( \iota^*\Omega \right)$ using the general theory reported in this paper is that it does not require to know the gauge transformations (equivalently, the parameters $e$ and $g$). The new formula (\ref{pdfc}) keeps this feature and, of course, it also works for point particle systems. 

To continue, let us recall some general facts about the Lagrangian formulation for field theory, which also helps us to fix the notation used throughout this paper. Let $\cal{C}$ be the configuration space of a field theory, which is usually assumed to be a differentiable Banach manifold whose points are labeled by $N$ field variables $\phi^A$, where $A$ denotes all possible indices that the field variables, which are functions, have. Thus, for instance, if the theory depends on differential forms then $\phi^A$ denotes their components. $T {\cal C}$ denotes the tangent space of ${\cal C}$, also called velocity phase space.  We discuss Lagrangians that depend on $\phi^A$ and its first derivatives only, as is usually the case (if the theory depends on higher derivatives it is possible to introduce auxiliary fields so that it becomes the mentioned type). The action principle is given by
\begin{eqnarray}
S[\phi^A]&=& \int_M \d^n x \, \mathcal{L}(\phi^A, \partial_{\mu} \phi^A),
\end{eqnarray}
with $M$ being the $n$-dimensional spacetime. Next, we assume spacetimes $M$ of the form $M= \mathbb{R}\times \Sigma$, and so
\begin{eqnarray}
S[\phi^A] &=& \int_{t_1}^{t_2} \d t \int_{\Sigma} \d^{n-1}x \, \mathcal{L}=: \int_{t_1}^{t_2} \d t \, L,
\end{eqnarray}
where $\mathcal{L}$ is the Lagrangian density, $L$ is called the Lagrangian of the theory, and $\partial_{\mu}:= \frac{\partial}{\partial x^{\mu}}$, where $(x^{\mu}):= (t,x^1, \ldots, x^{n-1}) \equiv (x^0, x^a)$ are spacetime local coordinates naturally adapted to the foliation $\mathbb{R} \times \Sigma$. Boundary terms are relevant in many aspects of the classic and quantum theory\cite{SunderSy} and, if they are present, a careful analysis must be done in order to have a well-defined action principle, for instance. However, boundary terms do not modify the equations of motion and so they do not change the number of physical degrees of freedom.  Then, we can avoid boundary terms either by taking $\Sigma$ without boundary or by imposing suitable boundary conditions on the fields (usually, that the fields vanish at spatial infinity). In what follows we restrict the analysis when $\Sigma$ has no boundary. In $\mathcal{L}$ we only write the variables that are dynamical, even when it can also depend upon non-dynamical background fields, such as the spacetime metric in special relativistic theories. Also, as usual, we only write $\phi^A$ in place of $\phi^A(x^{\mu})$. The equations of motion coming from this action principle are 
\begin{eqnarray}
\frac{\delta S}{\delta \phi^A}:= \frac{\partial \mathcal{L}}{\partial \phi^A}-\partial_{\mu} \left( \frac{\partial \mathcal{L}}{\partial \phi^A_{\mu}}\right)=0, \label{falta}
\end{eqnarray}
where $\phi^A_{\mu}:=\partial_{\mu} \phi^A$, and from now on we adopt Einstein's convention. In \eqref{falta} we have also defined the variational derivative $\frac{\delta}{\delta \phi^A}$. With the help of $\mathcal{L}$, it is possible to define a preferred symplectic two-form on $T {\cal C}$ \cite{woodhouse} written in local coordinates $\phi^A$ and ${\dot \phi}^A:=\partial_0 \phi^A$ as
\begin{eqnarray}
\Omega &=& \int \d^{n-1}x  \left[ \frac{\partial^2 {\mathcal L}}{ \partial \phi^A\partial \dot{\phi}^B} \, \ed \phi^A \ew \ed \phi^B \right.  + \frac{\partial^2 {\mathcal L}}{ \partial \phi^A_a \partial \dot \phi^B} \left( \partial_a \, \ed \phi^A \right)\ew \ed \phi^B \nonumber \\ 
&& \left.+ \frac{\partial^2 {\mathcal L}}{\partial \dot{\phi}^A \partial \dot \phi^B}\, \ed \dot \phi^A \ew \ed \phi^B \right],
 \label{gp1}
\end{eqnarray}
where we use $\ed$ and $\ewe$ to denote the infinite-dimensional exterior derivative and infinite-dimensional wedge product of forms on $T {\cal C}$, respectively (they are to be distinguished from $\d$ and $\wedge$ that are the finite-dimensional exterior derivative and finite-dimensional wedge product of forms on the spacetime, respectively). Since $\partial_{\mu}$ and $\ed$ act on different spaces, we have $\partial_{\mu}\, ( \ed \phi^A)= \ed (\partial_{\mu} \phi^A)$. Also, the notation $\int_{\Sigma_{t'}}$ instead of $\int \d^{n-1}x$ is sometimes used in literature. Notice that \eqref{gp1} is evaluated over a $\Sigma_{t}$ surface, this is the first difference from point particle systems: in that case, the symplectic structure is a sum, therefore it is not necessary to consider any particular surface. In this case, it seems natural to consider this surface because we speak of evolution of the fields (from one of these surfaces to another one) with respect to the parameter $t$, which is not necessarily a physical time. Also, this is in full analogy to Dirac's approach, where the Poisson bracket of two functions is calculated evaluating them at the same $t$. Note that this symplectic structure is different from the one used in the CCF (see Appendix \ref{apen2} for details).

The Lagrangian $L$ is said to be regular if $\Omega$ is non-degenerate, otherwise $L$ is singular or irregular. Notice that $\Omega$ is non-degenerate iff the Hessian $ W_{AB}=\frac{\partial^2 {\mathcal L}}{\partial \dot{\phi}^A \partial \dot \phi^B}$ is invertible (sometimes in the literature, when $\Omega$ is degenerate it is called presymplectic and not symplectic structure to emphasize its degeneracy; we will call it symplectic even in the degenerate case). Besides, in the singular case the Legendre map $FL: T { \cal C} \mapsto T^* {\cal C}$, is no longer invertible, i.e., there are functions on $T {\cal C}$ that cannot be projected to functions on the phase space $T^* {\cal C}$.

In  Subsections \ref{1A}-\ref{GT} we develop the necessary tools to analyze singular field theories from the Lagrangian viewpoint. We start by giving a brief summary of the GF based on Refs.~\onlinecite{Gotay, GotayNester}. It allows us to obtain the Lagrangian constraints of the theory under study using the constraint algorithm and the second-order equation problem which are conceptually equal to the case for point particles.

\subsection{Algorithm to obtain the Lagrangian constraints (constraint algorithm)}\label{1A}
If $\Omega$ is non-degenerate, we can write the Lagrangian equations of motion (\ref{falta}) as
\begin{equation}
X \cdot \Omega =- \,\ed E, \label{gp2+}
\end{equation}
where $X \cdot \Omega$ stands for the contraction of $X$ with $\Omega$ and 
\begin{eqnarray}\label{energy}
E(\phi^A, \dot{\phi}^A )= \int \d^{n-1} x \left(\dot{\phi}^A \frac{ \delta L}{ \delta \dot{\phi}^A } - \mathcal{L} \right)
\end{eqnarray}
is called the ``energy'' (even though it does not need to correspond to a notion of physical energy) and $X$ is completely determined by these equations. [Notice that the sign in (\ref{gp2+}) is different with respect to the one used in Ref.~\onlinecite{diaz}. This is because therein the symplectic structure was chosen as the negative of the analog of (\ref{gp1}) in point particles. Therefore, the difference of the global sign is just a matter of convention]. The differential of any function $f$ defined in $T {\cal C}$ is given by $\ed f= \int \d^{n-1}x \left( \frac{\delta f}{\delta \phi^A} \ed \phi^A+ \frac{\delta f}{\delta \dot{\phi}^A} \ed \dot{\phi}^A \right)$, which is frequently written as $\delta f= \int \d^{n-1}x\left(\frac{\delta f}{\delta \phi^A} \delta \phi^A+ \frac{\delta f}{\delta \dot{\phi}^A} \delta \dot{\phi}^A \right)$. From this perspective, the Lagrange equations (\ref{falta}) correspond to the integral curves of the vector field $X$. Moreover, when $\Omega$ is degenerate, we can still try to write the Lagrange equations (\ref{falta}) as the integral curves of a (to be determined) vector field $X=\int \d^{n-1}x \left( \alpha^A \frac{ \delta}{ \delta \phi^A}+ \beta^A \frac{ \delta}{ \delta \dot{\phi}^A} \right)$ on $T {\cal C}$ like in (\ref{gp2+}). In this case, Eqs. (\ref{gp2+}) become the Lagrange equations once we use $\alpha^A= \dot{\phi}^A $ and $\beta^A=\ddot{\phi}^A$ because this choice corresponds to the integral curves of $X$. However, in the singular case if $X$ satisfies (\ref{gp2+}) then $X+Z$ also does, where $Z$ is an arbitrary null vector of $\Omega$, and therefore $X$ is not unique. In addition, notice that there are points on the tangent bundle where $Z \cdot \ed E \neq 0$, which is inconsistent with (\ref{gp2+}). The inconsistency is solved by using the constraint algorithm described below. \cite{Gotay}

The constraint algorithm generates a sequence of sub-manifolds $T{\cal{C}} =: P_1 \supseteq P_2 \supseteq P_3 \supseteq \cdots$ which are defined by the Lagrangian constraints $\varphi's$ (many of them are actually part of the equations of motion). \cite{Gotay, diaz} The algorithm must end (if the theory is well--defined) at some final constraint submanifold $P := P_s \not = \emptyset $, $1\leq s < \infty$. 
Thereby, on $P$, we have completely consistent Lagrange's equations of motion
\begin{equation}
(X \cdot \Omega +\, \ed E)\mid_{P}= 0, \label{gp3}
\end{equation}
and at least one solution $X \in TP$ exists. However, the solutions may not be unique, and they are determined only up to the vector fields in $\ker \Omega\cap TP$.

Let us explain how to find the Lagrangian constraints. Consider the $ \ker \Omega$, if $Z\in  \ker \Omega$ then $Z \cdot \left( X \cdot \Omega\right)= \Omega(X,Z) =0$, and (\ref{gp2+}) requires that $Z \cdot \ed E =0$. Then, the points of $T\cal{C}$ where the equations (\ref{gp2+}) are inconsistent are those for which $Z \cdot \ed E \not=0$ for any $Z \in \ker \Omega$. Then $P_2$ is given by the points in $T {\cal{C}}$, which satisfy 
\begin{eqnarray}
\int \d^{n-1}x \, {\varphi_1}\textrm{'s} := \ker \Omega \cdot \ed E \approx 0 .
\end{eqnarray}

Now, we can try to solve
\begin{equation}
(X \cdot \Omega +\, \ed E)\mid_{P_2}= 0.
\end{equation}
This equation can be solved for $X$, but also, physically, we must demand that $X$ be tangent to $P_2$, i.e., we demand that the motion of the system takes place on $P_2$; this requirement is not always automatically accomplished, generating more Lagrangian constraints. This is the origin of $P_3, P_4$, etc. Namely, $X$ is tangent to $P_2$ iff $X \left( \int \d^{n-1}x \, {\varphi_1}\textrm{'s} \right)\approx 0$, where we must use that $X$ satisfies Eq. (\ref{gp2+}) and the constraints. This requirement could give rise to some constraints ${\varphi_2}$'s which define $P_3$, and we must also require $X$ to be tangent to $P_3$, and so on. Notice that some of the $\alpha$'s of $X$ are not fixed by Eq. (\ref{gp2+}) (because of the degeneracy of $\Omega$). Therefore, demanding that it be tangent to the Lagrangian constraint surface might yield new relationships involving these unknown components. If it is possible to eliminate them, by combining these relations, then we get a new set of Lagrangian constraints. We must demand $X$ be tangent to these new constraints, and so on. If it is not possible to eliminate the unknown components, those relations will also be Lagrangian constraints, once we use $\alpha^A=\dot{\phi}^A$ therein (see Subsection \ref{IIB}). This issue also arises in some of the point particle examples studied on Ref. ~\onlinecite{diaz} and for fields see the toy model in the Appendix \ref{apen1} of the present work. Finally, the constraint algorithm ends when the requirement that $X$ be tangent to the constraint surface does not give new Lagrangian constraints and/or only gives relations that involve the unknown components.

In this way, before we use $\alpha^A=\dot{\phi}^A$, we obtain $N_1+N_2-N^{(p)}$ Lagrangian constraints ($N^{(p)}$ is the number of primary constraints of the Hamiltonian analysis) that corresponds to the projectable ones, i.e., $FL (\varphi\textrm{'s})$ are all the secondary Hamiltonian constraints\cite{Gotay}  (see also Ref. \onlinecite{Batlle}).

\subsection{The second-order equation problem}\label{IIB}
Variational considerations require that the Lagrange equations (\ref{gp2+}) be a set of {\it second-order} differential equations. \cite{Nester} This requirement means
\begin{equation}\label{bogard}
X (\phi^A)= \dot{\phi}^A \Longleftrightarrow \alpha^A=\dot{\phi}^A.
\end{equation}
This condition and (\ref{gp2+}) generate more Lagrangian constraints [in the regular case, Eq. (\ref{gp2+}) always implies $\alpha^A=\dot{\phi}^A$, see Ref.~\onlinecite{GotayNester}] Also, as previously mentioned, this condition could be necessary in order to interpret the relations that could arise when we demand the vector $X$ be tangent to the constraint surface, giving more Lagrangian constraints (see appendix \ref{apen1}). The new Lagrangian constraints are the ($N^{(p)}-N^{(p)}_1$) non-projectable ones.\cite{Pons} Notice that we do not need to ask these constraints to be preserved over time. Actually, their time evolution give relations that involve accelerations, which are not Lagrangian constraints.\cite{Pons, Sundermeyer}
 
Summarizing, it is possible to obtain all the Lagrangian constraints $l= N_1 + N_2 - N^{(p)}_1$. Furthermore, we know which constraints are projectable and which are non-projectable. Also, in order to treat the constraints as functions in the tangent bundle, we need to smear them. So, we will work with $\varphi[M_s]:= \int \d^{n-1}x \, M_s \varphi^s$, where the index $s$ represents all the possible indices that the constraint could have, and $M_s=M_s(x^{\mu})$ are test functions.  

We recall that the non-geometrical approach to count the physical degrees of freedom of Ref.~\onlinecite{diaz} requires knowing $e$, $g$ and $l$. At this stage, we have $l$ only. On the other hand, $e$ and $g$ can be determined from the knowledge of the gauge transformations which can be obtained following different approaches; for instance the one described in Subsection \ref{GT} of this paper. Despite this, as we already mentioned, we find the geometric approach --which does not require to know the gauge transformations and so it does not require to know $e$ and $g$-- the natural way to deal with field theory. This geometric and Lagrangian approach is explained in Subsection \ref{IIC}. 

\subsection{Counting of the physical degrees of freedom}\label{IIC}
So far, we have identified the mathematical structure for field theory involved on the right-hand side of \eqref{gl1} and we have shown how we can obtain all the Lagrangian constraints, the question now is, how to calculate the rank of $\iota^*\Omega$. As we previously mentioned, it is not always possible to explicitly calculate $\iota^*\Omega$. This is the second difference that arises in this context, compared to the point particle case (it can be possible to use some of the constraints, only that in this case it will be necessary to recalculate the null vectors of the symplectic structure, and use those constraints in the expression for the energy $E$ to recalculate the new constraints. Therefore, it seems preferable not to use those constraints, unless all the projectable ones can be used). Now, we are going to explain the general procedure and then we show that, since we are working in the tangent space, we can take a shortcut for this calculation.  As a side remark, a similar development can be done in the Hamiltonian side. \cite{Teitelboim}

To compute the rank of $\iota^*\Omega$ we have to look for vector fields that satisfy (modulo surface terms)
\begin{equation}\label{vecf}
Z \cdot \Omega = \sum \ed \varphi[M_s],
\end{equation}
where $\sum$ means that this equation must be established for all possible linear combinations of the constraints (when this equation is established in the cotangent space, the corresponding vector fields are called {\it Hamiltonian vector fields}). So, first, we must establish  Eq. \eqref{vecf} for each constraint, then to take two of them in a linear combination, and so on. Notice that if $Z$ satisfies \eqref{vecf} then $Z+Z'$ also does, where $Z'$ is an arbitrary null vector of $\Omega$, and therefore $Z$ is not unique, i.e., it is determined up to the null vectors of $\Omega$. Also, from Eq. \eqref{vecf}, we have that the vector fields will depend on the parameters used to smear the constraints, i.e., $Z=Z[M^s]$. We call these vectors {\it associated vectors} because they are associated with the constraint(s) that appear in the right hand side of \eqref{vecf}. Under the inclusion map, they are candidates to be null vectors of the induced symplectic structure because the right-hand side of this equation vanishes, but not all of them are defined on the Lagrangian constraint surface. The vectors we are looking for are those tangent to the Lagrangian constraint surface. This means that they should override all the Lagrangian constraints, where we must take into account that these vectors are defined up to the null vectors of $\Omega$; i.e., they must satisfy
\begin{equation}\label{vtsc}
\left(Z+\ker \Omega \right) \left( \varphi[M_s] \right)\approx 0,
\end{equation}
for all the Lagrangian constraints. Not all vector fields that come from \eqref{vecf} satisfy this condition. We call {\it null associated vector fields} to those that do it.\cite{note1} Notice that in the linear combination $Z+\ker \Omega$, the vector fields of $\ker \Omega$ have their respective parameters, i.e., we ``smear" them (for example, if $Z_A=\int \d^{n-1} x \, \frac{\delta}{\delta \phi^A}$ belongs to $\ker \Omega$; the corresponding linear combination will be $Z_A[M^A]:=\int \d^{n-1} x \,M^A\frac{\delta}{\delta \phi^A} $, where $M^A$ is a smearing function), and the requirement that this vector field be tangent to the Lagrangian constraint surface fixes some of the parameters introduced in the vectors of $\ker \Omega$ to be functions of the parameters that appear in $Z$.
In fact, the condition \eqref{vtsc} gives a system of equations for the parameters of the vectors of $\ker \Omega$, whose solution, if it exists, can be obtained guided by covariance for covariance field theories, for instance. The parameters that are not fixed are those that smear vectors that are already tangent to the Lagrangian constraint surface. This means that the null associated vector fields are defined up to this kind of vector fields. Furthermore, in order to count the null associated vector fields, we will count how many independent parameters appear in their expressions. To calculate the rank of $\iota^*\Omega$, we must also take into account the independent constraints (or combinations of them) that appear on the right-hand side of (\ref{vecf}), i.e., how many independent times we can establish Eq. (\ref{vecf}). This is because these constraints reduce the rank of $\Omega$ (under the inclusion map).\cite{note2} Notice that this number coincides with the number of independent associated vector fields.

Before we calculate the rank of $\iota^*\Omega$ with the information we already have, two cases must be distinguished: the case where the constraints are independent and where they are not. They are called irreducible and reducible systems, respectively. For the irreducible case, the counting of the physical degrees of freedom is given by
\begin{eqnarray}
\frac{1}{2} {\rm Rank} \left ( \, \iota^*\Omega \right ) &=& N- \frac{1}{2}\left(B + C\right), \label{pdfc}
\end{eqnarray} 
where $B$ represents the number of independent null vectors (the original null vectors of $\Omega$ plus the null associated vector fields) and $C$ is the number of independent combinations of constraints involved in (\ref{vecf}) (which is the same as the number of associated vector fields). As the reader can suspect, in the reducible case, one must consider in (\ref{pdfc}) independent quantities only. We will explain in detail in Subsection \ref{IID} how this can be done.  Notice that we do not need to know the gauge transformations for the counting.

The procedure to count the null associated vector fields can be simplified thanks to the fact that we are working in the tangent space. We will use the following two facts: i) The first one is that $\Omega$ and the symplectic structure in the Hamiltonian formalism are related by the Legendre transformation. Actually, under the Legendre transformation, $\Omega$ becomes the standard symplectic structure on the cotangent space once it is restricted to the primary constraints.\cite{Gotay} Therefore, it only remains to restrict it to the secondary ones, which correspond to the projectable constraints in the Lagrangian formalism. Thus, only the projectable constraints can appear in the right-hand side of Eq. (\ref{vecf}). This means that for the non-projectable constraints there are no associated vectors. ii) The second fact, as we will see below, is that the null associated vector fields are essential in the construction of the gauge transformations. This fact is related with the second result, which is provided by the CCF. The mentioned result is that the gauge transformations are degenerate directions of the symplectic structure of the CCF {\it on-shell}.\cite{Wald} This means that the action of the symplectic structure of the CCF over those vectors is proportional to the equations of motion. Furthermore, we have shown in Appendix \ref{apen2} that the symplectic structure of the CCF looks like \eqref{gp1} if we consider a constant $t$ surface in that formalism, except that they both are defined on different manifolds and are endowed with different properties. Therefore, only the vector fields associated with the Lagrangian constraints that are part of the equations of motion will generate null associated vector fields. This is an easier way to identify (and count) them. Even so, these simplifications are useful if we only are interested in the counting of the degrees of freedom. If we want to compute the off-shell gauge transformations, the explicit form of the null associated vectors is needed. In the examples, we will apply the general procedure and show how it could be simplified by the considerations mentioned above.

\subsection{Reducibility}\label{IID}

Now, let us explain how we must deal with reducible systems. The separation between irreducible and reducible constraints is not always possible. Nonetheless, if the theory has reducibility this means that there are some $Z^r_K$ such that $Z^r_K \varphi_r=0$, where $\varphi_r$ are the reducible constraints and $K=1,2,...,k$, $k$ represents how many reducibility conditions exist (it can happen that $Z^r_K$ are differential operators, for example in the BF theory analyzed in Section \ref{sec3}). Then, if the $Z^r_K \varphi_r$ are irreducible, the number of irreducible constraints is $r-k$. This is the first stage of reducibility. However, the $Z^r_K \varphi_r$ could also be reducible; in such a case, they obey $Z^K _{K_1}Z^r_K \varphi_r=0$ for some $Z^K_{K_1}$ with $K_1=1,....,k_1$, $k_1$ being the number of these relations. This is called the second stage of reducibility; if $Z^K _{K_1}Z^r_K \varphi_r$ are irreducible, then we have $k-k_1$ independent reducibility conditions, and so $r-(k-k_1)=r-k+k_1$ irreducible constraints. The $Z^K _{K_1}Z^r_K \varphi_r$ relations of the second stage could also be reducible, giving rise to a third stage of reducibility, and so on. That is to say, for some theories there exist a finite number of stages of reducibility; for example, the $n$-dimensional BF theory possesses $n-3$ stages of reducibility in its Hamiltonian formulation \cite{Blau, Thompson} (see also Refs.~\onlinecite{Caicedo, Cuesta}). Our approach perfectly works in the case when there is a finite number of reducibility stages. There are other theories that have an infinite number of stages of reducibility;\cite{Beng,Gomis} this is not covered in our approach. In summary, if there are reducible constraints, we must count just the independent ones (like in Dirac's canonical formalism). 

However, in the Lagrangian approach, we not only count the Lagrangian constraints that appear in \eqref{vecf}, but also the null associated vector fields. Then, we need to know if the reducibility of the constraints has some implications on them. The answer is given by Eq. (\ref{vecf}). As we have already smeared the constraints, we can choose in \eqref{vecf} the smeared functions to be proportional to the $Z^r_K$. In that case, the right-hand side vanishes. Since the vectors that appear there are assumed to be different from the null vectors of $\Omega$ they must vanish too, at least over the Lagrangian constraint surface. This means that not all the parameters in the vector fields are irreducibly, and we must count just the irreducibly ones. Notice that if we have stages of reducibility in the constraints to which the vectors fields are associated we will have stages in the choice of the parameters.

Therefore, the procedure on the associated  and null associated vector fields consists of searching for a choice of the parameters such that some of the associated vector fields vanish over the Lagrangian constraint surface.  This is not a {\it brute force} procedure, because the reducibility conditions in the constraints determine the form of the parameters that override the associated vector fields. In summary, if there is reducibility, we have to count just independent quantities in \eqref{pdfc}. 

By using (\ref{gl1}) and (\ref{pdfc}), we have $e+g$ ($=B+C-l$). In spite of this, in order to finally identify the parameters $e$, and then $g$, we need the gauge transformations. In Subsection \ref{GT}, using the current information, we explain the steps that we must follow to obtain it. This show that the current geometric Lagrangian approach is robust enough to allow us to obtain this information too.

\subsection{Gauge transformations}\label{GT}
As is well known, the gauge transformations can be obtained following different approaches; for example, using the converse of Noether's second theorem \cite{Noether,translation} (see Ref.~\onlinecite{Celada} where this approach was used to reveal the gauge symmetries of first-order general relativity. The same approach can be applied to theories analyzed in Sec. \ref{sec3}). It is worth mentioning that we can also obtain the gauge transformations with the information we already have.  In Ref.~\onlinecite{diaz} it was shown that the geometric Lagrangian approach allows us to obtain the {\it on-shell} gauge transformations. In this subsection, we extend this result and show how to construct the {\it off-shell} gauge transformations by following two steps:

i) First, we need the null associated vector fields and the vector fields of $\ker \Omega$ that are tangent to the Lagrangian constraint surface (those whose parameters have not been fixed when we search for the null associated vectors), which we have already found to make the counting of the physical degrees of freedom. With them, we form a unique vector $\tilde{X}$ as a linear combination.
Notice that the smeared functions that appear in this vector field are actually the gauge parameters of the gauge transformations. 
 
ii) Second, in order for this vector to represent a vector field whose components are the gauge transformations, it must be of the form $\tilde{X}= \int \d^{n-1} x\left[ \delta_\varepsilon \phi^A \frac{\delta}{\delta \phi^A}+ \partial_0(\delta_{\varepsilon} \phi^A) \frac{\delta}{\delta \dot{\phi}^A} \right]$, which implies that $ \delta_{\varepsilon} \dot{\phi}^A= \partial_0(\delta_{\varepsilon} \phi^A)$. This is because, as is well known, even when we consider $\phi^A$ and $\dot{\phi}^A$ as independent variables, their variations are not. This gives some relations between the gauge parameters, which completely determine the gauge transformations [we emphasize that in the geometric approach followed in Ref.~\onlinecite{diaz} we explicitly calculated 
$\iota^*(\tilde{X} \cdot \Omega)$, which gives the gauge transformations on-shell. We recall that off-shell and on-shell gauge transformations do not always coincide]. 

Once we have the gauge transformations we can count $e$, and $g$. If there is reducibility, we must be careful in the counting of the parameters because as we have already shown reducibility is also reflected in the gauge parameters, and if some of them are not independent, their time derivatives are also not independent. With this information, we can use the map (\ref{map}) and calculate--if we wished--how many independent first- and second-class constraints follow from Dirac's canonical analysis of the theory without having to perform any canonical analysis.

On the other hand, as we previously mentioned, at the end we have obtained  all the Lagrangian parameters $e, g$, and $l$, and we could use the formula \eqref{gl1} for the counting if we want. Despite that, as we have remarked, the use of the formula \eqref{pdfc} has the advantage that can be used {\it without} knowing the gauge transformations.

To summarize, in this section, we have shown that it is possible to obtain all the Lagrangian constraints, and the off-shell gauge transformations (if any) for arbitrary singular field theories in a geometric way. We have also presented the steps that we must follow to make the counting of the physical degrees of freedom, which is the main result of this paper.

%\\\\\\\\\\\\\\\\\Section III\\\\\\\\\\\\\\\\\\\\\\
\section{Examples}\label{sec3}
In this section, we illustrate the theoretical framework developed in Sec. \ref{sec2} in several physical field theories. In particular, we analyze the three-dimensional Palatini action with and without a cosmological constant, Witten's exotic action, three-dimensional Chern-Simons action, and BF theory in four dimensions (which possesses reducibility conditions). In all these theories the internal gauge group is an arbitrary Lie group. We also study the first-order general relativity in four dimensions described by Palatini's action with a cosmological constant. 

Let us establish here some notation used in this section. We denote by $G$ the semi-simple Lie group of the theory. We call $\mathfrak{g}$ the Lie algebra of $G$, we denote its structure constants by $f^i{}_{jk}$, and by $k_{ij}$ its non-degenerate Killing form, which is used to raise and lower the internal indices $i,j,k, \dots $, which run from 1 to dim $\mathfrak{g}$. We use $\mu ,\nu ,\rho,...$, which run from $0$ to $n-1$ for spacetime indices and $a,b,c, \ldots =1, \ldots ,(n-1)$ for space. For $\epsilon^{\mu_0 \mu_1 \cdots \mu_{(n-1)}}$ we take $\epsilon^{012 \cdots (n-1)}=1$. Remember we use $\partial_{\mu}:= \frac{\partial}{\partial x^{\mu}}$, where $x^{\mu}:=\{ t,x^a\}$ are the spacetime coordinates, and the dot over the variable means the action of $\partial_0$ thereon. Finally, symmetrization and antisymmetrization of indices are denoted by brackets, according to $A^{(\mu \nu)}:=\frac{1}{2}\left( A^{\mu \nu} + A^{\nu\mu} \right)$, and $A^{[\mu \nu]}:=\frac{1}{2}\left( A^{\mu \nu} - A^{\nu\mu } \right)$, respectively. 

%\\\\\\\\\\\\\\\\\\\\\\\\\\\\\\\\\\\\\\\\\\\\\\\\\\\\\\\\\\\\\\\\\\\\\\\\
\subsection{Three-dimensional generalized Palatini theory with $\Lambda \neq 0$}\label{ejem1}
Let us consider a generalized three-dimensional Palatini action based on $G$ with cosmological constant $\Lambda$. The term ``generalized'' means that we are considering an arbitrary Lie group.\cite{Romano} This action is also known as the action for a three-dimensional BF theory with cosmological constant $\Lambda$.\cite{Review} For the particular Lie groups $SO(2,1)$ and $SO(3)$, this theory is three-dimensional general relativity with a cosmological constant for Lorentzian and Euclidean signatures, respectively. We assume, as in Ref.~\onlinecite{Romano}, that $G$ admits a totally anti-symmetric invariant tensor $\epsilon^{ijk}$, which satisfies $\epsilon^{i[jk}f^{l]}{}_{im}=0$ (a possible choice for $\epsilon^{ijk}$ is given by $\epsilon^{ijk}= f^i{}_{lm}k^{lj}k^{mk}$, this is not the only one\cite{Romano}). The action is given by 
\begin{eqnarray}
S[e,A]&=&\int_M \left( e_i \wedge F^i - \frac{\Lambda}{6}\epsilon^{ijk} e_i \wedge e_j \wedge e_k \right), \label{palati} 
\end{eqnarray}
where $e^i=e^i_{\mu} \d x^{\mu}$ is a $\mathfrak{g}$-valued $1$-form, $A^i=A^i_{\mu} \d x^{\mu}$ is a $\mathfrak{g}$-valued connection $1$-form and $F^i = \d A^i + \frac{1}{2} f^i\,_{jk} A^j \wedge A^k = \frac{1}{2} F^i_{\mu \nu} \d x^{\mu}\wedge \d x^{\nu}$ its the curvature, $F^i_{\mu \nu}=2\partial_{[\mu} A^i_{\nu]}+f^i{}_{jk} A^j_{\mu}A^k_{\nu}$. Note that both the Lie algebra index $i$ and the spacetime index $\mu$ correspond to the index $A$ of the general theory in Sec. \ref{sec2}. Therefore, we have $N=(3 + 3) \times{\rm dim} \, \mathfrak{g}$ configuration variables in $e^i_{\mu}$ and $A^i_{\mu}$. The variation of the action principle (\ref{palati}) with respect to the independent variables yields the variational derivatives ${\mathcal E}^i$ and ${\mathcal E}'^i$ 
\begin{eqnarray}
\delta e^i &:&  {\mathcal E}^i := F^i-\frac{\Lambda}{2} \epsilon^{ijk}e_{j} \wedge e_{k},
% \nonumber\\ &&\Leftrightarrow\epsilon^{\mu\nu\rho} \left( F^i_{\nu\rho}-\Lambda \epsilon^{ijk}e_{j \nu} e_{k \rho} \right)=0,
\nonumber \\
\delta A^i &:& {\mathcal E}'^i:= \mathcal{D}e^i, %\Leftrightarrow \epsilon^{\mu\nu\rho} \mathcal{D}_{\nu}e^i_{\rho} =0 
\label{imp2}
\end{eqnarray}
where $\mathcal{D}v^i:= \d v^i+f^i{}_{jk} A^j \wedge v^k$ is the covariant derivative of the $\mathfrak{g}$-valued $p$-form $v^i$. So, the equations of motion are
\begin{eqnarray}
{\mathcal E}^i =0, \qquad {\mathcal E}'^i=0. \label{empa}
\end{eqnarray}

By performing the $2 + 1$ decomposition of the fields, the action (\ref{palati}) acquires the form
\begin{eqnarray}\label{palati2+1}
S[e, A] &=& \int_M \d^3x \, \epsilon^{0ab}  \left[-e_{ia} \dot{A}^i_b +e_{ia}\mathcal{D}_b A^i_0 
%\right. \nonumber\\ &&\left. 
+\frac{e_{i0}}{2}\left( F^i_{ab}  %\right. \right. \nonumber\\ &&\left. \left. 
-\Lambda\epsilon^{ijk}e_{ja}e_{kb}\right) \right].
\end{eqnarray} 
{\it Lagrangian constraints}. The points of the tangent bundle $T {\mathcal C}$ are locally labeled with $12 \times{\rm dim} \, \mathfrak{g}$ variables given by $(e^i_{\mu}, A^i_{\mu}, {\dot e}^i_{\mu}, {\dot A}^i_{\mu})$.  From (\ref{palati2+1})--modulo surface terms--the corresponding symplectic structure (\ref{gp1}), energy (\ref{energy}), and differential of the energy are
\begin{eqnarray}\label{psp}
\Omega &=&-\int \d^{2}x \, \epsilon^{0ab} \ed e_{ia} \ew \ed A^i_b, \nonumber\\
E &=&\int \d^{2}x \, \epsilon^{0ab} \left[A_{i0} \mathcal{D}_b e^i_{a} -\frac{e_{i0}}{2}\left( F^i_{ab} 
%\right. \right. \nonumber\\ && \left. \left. 
-\Lambda\epsilon^{ijk}e_{ja}e_{kb}  \right) \right], \nonumber \\
%\ed E&=& \int d^{2}x\epsilon^{0ab}\left[ -\frac{1}{2}\left( F^i_{ab}-\Lambda\epsilon^{ijk}e_{ja}e_{kb}\right) \ed e_{i0} \right. \nonumber\\
%&& + \left(\mathcal{D}_a A^i_0+ \Lambda \epsilon^{ijk} e_{j0}e_{ka}\right)\ed e_{ib} +\left( \mathcal{D}_a e^i_{0} \right.\nonumber \\
%&& \left.\left. -f^i{}_{jk} A^j_0e^k_{a} \right)\ed A_{ib}  +\mathcal{D}_be^i_{a}\, \ed A_{i0} \right].\\
\ed E&=& \int \d^{2}x\, \epsilon^{0ab}\left[ \mathcal{D}_be^i_{a}\, \ed A_{i0}+\left( \mathcal{D}_a e^i_{0}  -f^i{}_{jk} A^j_0e^k_{a} \right)\ed A_{ib}   \right. \nonumber\\
&&-\frac{1}{2}\left( F^i_{ab}-\Lambda\epsilon^{ijk}e_{ja}e_{kb}\right) \ed e_{i0}  + \left(\mathcal{D}_a A^i_0  
%\right. \nonumber\\&&\left. 
\left. + \Lambda \epsilon^{ijk} e_{j0}e_{ka}\right)\ed e_{ib} \right].
\end{eqnarray} 
Using this information, we look for the vector field $X$ in (\ref{gp2+}),
\begin{equation}
X:=\int \d^{2}x \left( \alpha^i_ {\mu} \frac{\delta}{\delta e^i_{\mu} }+\alpha'^i_{\mu} \frac{\delta}{\delta A^i_{\mu} } +\beta^i_{\mu} \frac{\delta}{\delta\dot{e}^i_{\mu} }+\beta'^i_{\mu} \frac{\delta}{\delta \dot{A}^i_{\mu} }\right),
\end{equation}
and we get
\begin{eqnarray}
\epsilon^{0ab}\alpha^i_{a} &=& \epsilon^{0ab}\left( \mathcal{D}_a e^i_{0}-f^i{}_{jk} A^j_0e^k_{a} \right), \nonumber\\
\epsilon^{0ab} \alpha'^i_{a}&=& \epsilon^{0ab} \left( \mathcal{D}_a A^i_0+ \Lambda \epsilon^{ijk} e_{j0}e_{ka}\right), \nonumber\\
 0&=& \epsilon^{0ab} \left( F^i_{ab}-\Lambda\epsilon^{ijk}e_{ja}e_{kb}\right),\nonumber \\
 0&=&-\epsilon^{0ab} \mathcal{D}_b e^i_{a},\label{eapalg}
\end{eqnarray}
which agree with (\ref{imp2}) once we substitute $\alpha^i_{a}= \dot{e}^i_{a}$ and $\alpha'^i_{a}= \dot{A}^i_{a}$ in the LHS of the first two lines of (\ref{eapalg}). 

It follows that $\operatorname{Rank} \Omega = 4 \times{\rm dim} \, \mathfrak{g}$ and that a basis of $\ker \Omega$ is given by the $8 \times {\rm dim}\, \mathfrak{g}$ vector fields $\{ Z_{1i}, Z_{2i}, Z_{3 i}^{\mu}, Z_{4 i}^{\mu} \}$,
\begin{eqnarray}\label{inv}
Z_{1i} &:=& \int \d^{2}x \, \frac{\delta}{\delta e^i_0}, \qquad
Z_{2i} := \int \d^{2}x \, \frac{\delta}{\delta A^i_0} ,\nonumber\\
Z_{3i}^{\mu} &:=& \int \d^{2}x \, \frac{\delta}{\delta \dot{e}^i_{\mu}} , \qquad Z_{4 i}^{\mu} := \int \d^{2}x \, \frac{\delta}{\delta \dot{A}^i_{\mu}}.
\end{eqnarray}
Nevertheless, among the elements of this set, only $ Z_{1i}$ and $Z_{2i}$ generate Lagrangian constraints, given by
\begin{eqnarray}\label{etiqueta}
\int \d^2x \, \varphi_{1i} & := & Z_{1i}\cdot \ed E=-\frac{1}{2} \int \d^2x \, \epsilon^{0ab} \left( F_{iab}-\Lambda\epsilon_{ijk}e^j_{a}e^k_{b}\right) \approx 0, \nonumber\\
\int \d^2x \, \varphi_{2i} & :=& Z_{2i}\cdot \ed E= \int \d^2x \, \epsilon^{0ab} \mathcal{D}_be_{ia} \approx 0.
\end{eqnarray}
Notice that these $(1+1) \times {\rm dim} \, \mathfrak{g} =2 \times {\rm dim} \, \mathfrak{g}$ constraints are part of the equations of motion (\ref{imp2}). Continuing with the approach, we must demand that $X\left( \int \d^2x \, \varphi_{1i} \right) \approx 0 \approx X \left( \int \d^2x \, \varphi_{2i}\right)$. Since $X$ satisfies (\ref{eapalg}), and using the Lagrangian constraints (\ref{etiqueta}), this is automatically fulfilled. Therefore, the constraint algorithm gives us $2 \times{\rm dim} \, \mathfrak{g}$ {\it projectable} Lagrangian constraints (applying the Legendre transformation to them yields the usual secondary constraints in Dirac's canonical analysis\cite{Romano}).

Next, we look for non-projectable Lagrangian constraints. The requirements $\alpha^i_{a}= \dot{e}^i_{a}$ and $\alpha'^i_{a}= \dot{A}^i_{a}$ in (\ref{eapalg}) imply that the fields must satisfy the $ (2+2) \times {\rm dim} \, \mathfrak{g} = 4 \times {\rm dim} \, \mathfrak{g}$ {\it non-projectable} Lagrangian constraints,
\begin{eqnarray}\label{cnp3d}
\varphi_{1}^{ia} & := & \epsilon^{a\mu \nu} \left( F^i_{\mu \nu}-\Lambda \epsilon^i{}_{jk} e^j_{\mu} e^k_{\nu} \right)\approx 0, \nonumber\\
\varphi_{2}^{ia} & :=& \epsilon^{a \mu \nu} \mathcal{D}_{\mu} e^i_{\nu}\approx 0.
\end{eqnarray}
Therefore, the total number of (projectable and non-projectable) Lagrangian constraints is $l= (2+4) \times{\rm dim} \, \mathfrak{g} = 6 \times{\rm dim} \, \mathfrak{g}$.% Notice that, for this field theory, all the equations of motion (\ref{imp2}) are Lagrangian constraints.

\subsubsection{Degree of freedom count}\label{IIIAI}
{\it Associated vector fields}. Due to the fact the variables $A^i_0$ and $e^i_0$ appear in the non-projectable constraints (\ref{cnp3d}), and that they do not appear in the symplectic structure $\Omega$ in (\ref{psp}), it follows from (\ref{vecf}) that there are no associated vector fields linked to the non-projectable constraints (\ref{cnp3d}), in full agreement with the general theory of Sec. \ref{sec2}. With respect to the projectable constraints (\ref{etiqueta}), we smear them with test fields $N^i$ and $M^i$ 
\begin{equation}\label{scp}
\varphi_1[N^i]:=\int \d^2x \, N^i \varphi_{1i}, \qquad \varphi_2[M^i]:=\int \d^2 x \, M^i \varphi_{2i},
\end{equation}
and from (\ref{vecf}) it can readily be seen that their associated vector fields--modulo the null vector fields (\ref{inv})--are
\begin{eqnarray}
Z_5[N^i]&=& -\int \d^2x \left( \mathcal{D}_a N^i \frac{\delta}{\delta e^i_a}- 
 \Lambda \epsilon^i{}_{jk} e^j_a N^k \frac{\delta}{\delta A^i_a} \right), \nonumber\\
Z_6 [M^i]&=& -\int \d^2x \left( f^i{}_{jk} e^j_a M^k \frac{\delta}{\delta e^i_a}+ \mathcal{D}_a M^i \frac{\delta}{\delta A^i_a} \right). \label{nullvecpa}
\end{eqnarray}
In fact, modulo a surface integral, they satisfy $Z_5[N^i] \cdot \Omega = \ed \varphi_1[N^i] $ and $Z_6 [M^i] \cdot \Omega=\, \ed \varphi_2[M^i]$. As we mentioned in Sec. \ref{sec2}, these vectors depend on the parameters used to smear the constraints and there are $2 \times{\rm dim} \, \mathfrak{g}$ of them.

{\it Null associated vector fields}. Notice that the associated vector fields (\ref{nullvecpa}) are associated with Lagrangian constraints that are part of the equations of motion (\ref{imp2}), therefore we expect that they are indeed null associated vector fields. If we used that criterion, we could make the counting of the physical degrees of freedom right now. Nonetheless, instead of that, we continue with the general procedure and we have to show that they are indeed null associated vector fields, i.e., we have to check if there are some linear combinations of these vectors and \eqref{inv} such that they are tangent to the Lagrangian constraint surface defined by \eqref{etiqueta} and \eqref{cnp3d}. Thus, we add to each of the associated vector fields (\ref{nullvecpa}) an arbitrary linear combination of the original null vectors (\ref{inv}) (the adopted convention is, for example, $Z_{1i}[P^{1i}]:=\int \d^2 x \, P^{1i}\frac{\delta}{\delta e^i_0}$. We recall that $Z_{1i}=\int \d^2 x \, \frac{\delta}{\delta e^i_0}$),
\begin{eqnarray}
Z_7&:=& Z_5\left[N^i\right]+ Z_{1i}\left[P^{1i}\right]+ Z_{2i}\left[P^{2i}\right]+Z_{3i}^{\mu}\left[ P^{3i}_{\mu} \right]+Z_{4 i}^{\mu}\left[ P^{4i}_{\mu}\right], \nonumber \\
Z_8&:=& Z_6\left[M^i\right]+ Z_{1i}\left[Q^{1i}\right]+ Z_{2i}\left[Q^{2i}\right]+Z_{3i}^{\mu}\left[ Q^{3i}_{\mu} \right]+Z_{4 i}^{\mu}\left[Q^{4i}_{\mu}\right], \label{nvfp}
\end{eqnarray}
and we demand the vector fields $Z_7$ and $Z_8$ to be tangent to the Lagrangian constraint surface 
$\varphi_{1i} \approx 0$, $\varphi_{2i} \approx 0$, $\varphi^{a}_{1i} \approx 0$, and $\varphi^{ia}_{2} \approx 0$. They are indeed tangent vectors iff they acting on all the smeared (projectable and non-projectable) Lagrangian constraints $\varphi [\mbox{smearing fields}]$ satisfy 
\begin{eqnarray}\label{parara}
Z_7 \, \left ( \varphi [\mbox{smearing fields}] \right ) \approx 0, \qquad  Z_8\,  \left ( \varphi [ \mbox{smearing fields}] \right ) \approx 0.
\end{eqnarray}
[cf. (\ref{vtsc})]. Therefore, computing the action of $Z_7$ and $Z_8$ on the smeared Lagrangian constraints we have
\begin{eqnarray}
Z_7 \, \left ( \varphi_1[N^i] \right ) &=& 0, \label{ne1} \\
Z_7 \, \left ( \varphi_2[M^i] \right ) &=& \int \d^2 x \, \left[ - f^i{}_{jk}N^i M^j \varphi^k_1 \right] \approx 0, \label{ne2}\\
Z_7 \, \left ( \varphi_1[N_{ia}] \right ) &=& Z_7 \, \left ( \int \d^2x \, N_{ia} \, \varphi^{ia}_{1} \right ) \nonumber\\
&=& \int \d^2 x \, 2N_{ia} \epsilon^{0ab}\left[ \mathcal{D}_b P^{2i}-P^{4i}_b+f^i{}_{jk}\left(  \Lambda \epsilon^j{}_{lm}e^l_b N^m \right) A^k_0 \right. \nonumber\\
&& \left.-\Lambda\epsilon^i{}_{jk}\left(e^j_b P^{1k}- e^k_0\mathcal{D}_b N^j \right)   \right] \nonumber\\
&=& \int \d^2 x \, 2N_{ia} \left( \epsilon^{0ab}\left\{ \mathcal{D}_b \left[ P^{2i}-\Lambda \epsilon^i{}_{jk} e^j_0 N^k \right] - \Lambda\epsilon^i{}_{jk} e^j_b \left[P^{1k}+\mathcal{D}_0 N^k\right]  \right. \right. \nonumber\\
&&\left. \left. -\left[ P^{4i}_b-\partial_0 \left( \Lambda \epsilon^i{}_{jk} e^j_b N^k\right) \right] \right\} + \Lambda \epsilon^i{}_{jk}\varphi^{ja}_2 N^k \right), \label{ne3}\\
Z_7 \, \left ( \varphi_2 [M_{ia}] \right ) &=& Z_7 \, \left ( \int \d^2 x \, M_{ia} \, \varphi_{2}^{ia} \right ) \nonumber\\
&=& \int \d^2 x \, M_{ia} \epsilon^{0ab} \left[ \mathcal{D}_b P^{1i}-P^{3i}_b+f^i{}_{jk} \left( \Lambda e^k_0\epsilon^{j}{}_{lm}e^l_b N^m -P^{2j}e^k_b+A^j_0 \mathcal{D}_b N^k \right)  \right]\nonumber \\
 &=& \int \d^2 x \,  M_{ia} \left( \epsilon^{0ab}\left\{ \mathcal{D}_b \left[ P^{1i}+ \mathcal{D}_0 N^i \right] -f^i{}_{jk} e^j_b \left[P^{2k}-\Lambda \epsilon^k{}_{lm} e^l_0 N^m\right] \right. \right. \nonumber\\
 &&\left. \left. -\left[ P^{3i}_b+\partial_0 ( \mathcal{D}_b N^i) \right]   \right\}  + \frac{1}{2} f^i{}_{jk} \varphi^{aj}_1 N^k \right), \label{ne4}\\
Z_8 \, \left ( \varphi_1[N^i] \right ) &=& \int \d^2 x \, \left[ f^i{}_{jk}N^i M^j \varphi^k_1   \right] \approx 0, \label{ne5}\\
Z_8 \, \left ( \varphi_2[M^i] \right ) &=& 0, \label{ne6}\\
Z_8 \, \left ( \varphi_1[N_{ia}] \right ) &=&\int \d^2 x \, 2N_{ia} \epsilon^{0ab} \left[ \mathcal{D}_b Q^{2i}-Q^{4i}_b-f^i{}_{jk}A^k_0\mathcal{D}_b M^j \right. \nonumber\\
&&\left.-\Lambda\epsilon^i{}_{jk}\left(e^j_b Q^{1k}-f^{j}{}_{lm}e^l_b M^m e^k_0\right)  \right] \nonumber\\
&=& \int \d^2 x \, N_{ia}\left( 2\epsilon^{0ab}\left\{  \mathcal{D}_b \left[ Q^{2i}-\mathcal{D}_0 M^i \right]- \Lambda\epsilon^i{}_{jk}e^j_b \left[ Q^{1k}+f^k{}_{lm} e^l_0 M^m\right] \right. \right. \nonumber \\ 
&& -\left. \left.\left[ Q^{4i}_b+\partial_0 \left( \mathcal{D}_b M^i \right) \right] \right\} + f^i{}_{jk}\varphi^{ja}_1 M^k \right), \label{ne7}\\
Z_8 \, \left ( \varphi_2[M_{ia}] \right ) &=&\int \d^2 x \, M_{ia} \epsilon^{0ab} \left[ \mathcal{D}_b Q^{1i}-Q^{3i}_b-f^i{}_{jk} \left( e^k_0 \mathcal{D}_b M^j+Q^{2j}e^k_b-A^j_0 f^k{}_{lm}e^l_b M^m \right)  \right] \nonumber \\
 &=& \int \d^2 x \,  M_{ia} \left( \epsilon^{0ab}\left\{ \mathcal{D}_b \left[ Q^{1i}+f^i{}_{jk} e^j_0 M^k \right] +f^i{}_{jk} e^j_0 \left[Q^{2k}+\mathcal{D}_0 M^k\right] \right. \right. \nonumber \\
 &&\left. \left. -\left[ Q^{3i}_b+\partial_0 (f^i{}_{jk} e^j_b M^k) \right]   \right\}  +f^i{}_{jk} \varphi^{aj}_2 M^k \right).  \label{ne8}
\end{eqnarray}
Notice that (\ref{ne1}) and (\ref{ne6}) are strongly zero, while (\ref{ne2}) and (\ref{ne5}) are weakly zero (they vanish on the Lagrangian constraints).  We have also dropped out the surface terms in \eqref{ne3}, \eqref{ne4}, \eqref{ne7}, and \eqref{ne8}. Requiring \eqref{ne3} and \eqref{ne4} vanish on the Lagrangian constraint surface implies
\begin{align}
&P^{1i} \approx -\mathcal{D}_0 N^i , &&P^{2i}\approx\Lambda \epsilon^i{}_{jk} e^j_0 N^k, \nonumber \\
&P^{3i}_{a}\approx -\partial_0 \left( \mathcal{D}_a N^i\right), &&P^{4i}_{a}\approx\partial_0 \left(\Lambda \epsilon^i{}_{jk} e^j_a N^k \right), \label{separada}
\end{align}
while requiring \eqref{ne7} and \eqref{ne8} vanish on the Lagrangian constraint surface implies
\begin{align}
&Q^{1i}\approx - f^i{}_{jk} e^j_{0} M^k , &&Q^{2i}\approx -\mathcal{D}_{0} M^i, \nonumber \\
&Q^{3i}_{a}\approx -\partial_0\left( f^i{}_{jk} e^j_{a} M^k \right), &&Q^{4i}_{a}\approx -\partial_0\left( \mathcal{D}_{a} M^i\right). \label{faltaesta}
\end{align}
The parameters $P^{3i}_{0}$, $P^{4i}_{0}$, $Q^{3i}_{0},$ and $Q^{4i}_{0}$ are not fixed and so they are left arbitrary because they smear the vector fields $ Z_{3i}^{0}$ and $Z_{4i}^{0}$, which are tangent to the Lagrangian constraint surface.

The form of the parameters in \eqref{separada} and \eqref{faltaesta} is already expected. In fact, the parameters $P^{1i},P^{2i}, Q^{1i}$ and $Q^{2i}$ could have been obtained by covariance too. For example $P^{1i}$ are the components along the vector fields associated with $e^i_0$, and from $Z_5$ we know that the components associated with $e^i_a$ are $-\mathcal{D}_a N^i$, and then by covariance it is expected that $P^{1i}=-\mathcal{D}_0 N^i$. On the other hand, as we have explained in the general theory, the null associated vectors are useful to find the gauge transformations and they have the property that the form of the components of the vector field along the $\dot{\phi}^A$ must be the dot of the components along the $\phi^A$. For example, if $P^{3i}_{a}$ are the components along $\dot{e}^i_a$, then they must be the dot of the components along $e^i_a$, and we know form $Z_5$ that they are $-\mathcal{D}_a N^i$, then $P^{3i}_{a}= -\partial_0(-\mathcal{D}_a N^i)$. This is in full agreement with the results that we have obtained from the system \eqref{parara}. In practice, we can use this reasoning to propose the expression for the parameters.

Therefore, as we anticipated, the $(1+1) \times{\rm dim} \, \mathfrak{g}= 2\times{\rm dim} \, \mathfrak{g}$ vectors \eqref{nullvecpa} become null associated vector fields, and because of  (\ref{nvfp}) and \eqref{faltaesta} they are given by
\begin{eqnarray}
Z_7[N^i]&=& -\int \d^2x \left[ \left( \mathcal{D}_{\mu} N^i \right)\frac{\delta}{\delta e^i_{\mu}} + \left( - \Lambda \epsilon^i{}_{jk} e^j_{\mu} N^k \right) \frac{\delta}{\delta A^i_{\mu}} + \partial_0 \left( \mathcal{D}_{a} N^i \right) \frac{\delta}{\delta \dot{e}^i_{a}} \right.\nonumber\\
&& \left.  + \partial_0 \left(- \Lambda \epsilon^i{}_{jk} e^j_a N^k \right) \frac{\delta}{\delta \dot{A}^i_{a}} \right], \nonumber\\
Z_8 [M^i]&=& -\int \d^2x \left[ f^i{}_{jk} e^j_{\mu} M^k \frac{\delta}{\delta e^i_{\mu}}+ \mathcal{D}_{\mu} M^i \frac{\delta}{\delta A^i_{\mu}} + \partial_0\left( f^i{}_{jk} e^j_{a} M^k \right) \frac{\delta}{\delta \dot{e}^i_{a}}\right.\nonumber\\
&& \left.  + \partial_0\left( \mathcal{D}_{a} M^i \right) \frac{\delta}{\delta \dot{A}^i_{a}} \right]. \label{nvfp_missing}
\end{eqnarray}

{\it Degree of freedom count}. We are ready to make the counting of the physical degrees of freedom: the number of field variables $e^i_{\mu}$ and $A^i_{\mu}$ is $N=6 \times{\rm dim} \, \mathfrak{g}$, the total number of null vectors, given by (\ref{inv}) and (\ref{nvfp_missing}), is $B=(8+2) \times{\rm dim} \, \mathfrak{g}= 10 \times{\rm dim} \, \mathfrak{g}$, and the number of linearly independent combinations of constraints or, equivalently, the number of associated vector fields in (\ref{nullvecpa}) is $C=2 \times{\rm dim} \, \mathfrak{g}$. Therefore, using \eqref{pdfc}, the number of physical degrees of freedom is
\begin{equation}
\frac{1}{2} {\rm Rank} \, \iota^*\Omega = \left[6-\frac{1}{2} \left(10+2 \right) \right]\times {\rm dim} \, \mathfrak{g}=0.
\end{equation}
Furthermore, by substituting this result into the right-hand side of Eq. (\ref{gl1}) we get $g+e=2N - l = \left[2 (6)- 6 \right] \times \mathfrak{g} = 6 \times{\rm dim} \, \mathfrak{g}$, which is the number of independent null vectors of $\iota^* \Omega$.

\subsubsection{Off-shell gauge transformations} \label{IIIAII}
We now look for the off-shell gauge transformations. Following the steps explained in \ref{GT}, first we use only the null associated vectors $Z_7$ and $Z_8$ given in Eq. (\ref{nvfp_missing}) and the null vector fields, $Z_{3i}^{0}$ and $Z_{4i}^{0}$, that are tangent to the Lagrangian constraint surface. With them, we form the vector field 
\begin{eqnarray}
\tilde{X}&:=& Z_7[-\rho^i]+Z_8 [-\tau^i]+Z_{3i}^{0}[\varepsilon^i_{30}]+Z_{4 i}^{0}[\varepsilon^i_{40}]\nonumber\\
&=& \int \d^2x \left[\left( \mathcal{D}_{\mu} \rho^i+f^i{}_{jk} e^j_a\tau^k \right)\frac{\delta}{\delta e^i_{\mu}}  +\left( \mathcal{D}_{\mu} \tau^i  - \Lambda \epsilon^i{}_{jk} e^j_{\mu} \rho^k \right)\frac{\delta}{\delta A^i_{\mu}} \right.\nonumber \\
&& + \partial_0 \left( \mathcal{D}_{a} \rho^i+f^i{}_{jk} e^j_{a} \tau^k \right) \frac{\delta}{\delta \dot{e}^i_{a}} + \partial_0 \left( \mathcal{D}_{a}\tau^i - \Lambda \epsilon^i{}_{jk} e^j_{a}\rho^k \right) \frac{\delta}{\delta \dot{A}^i_{a}}\nonumber\\
&& \left.+\varepsilon^i_{30} \frac{\delta}{\delta \dot{e}^i_{0}}+\varepsilon^i_{40} \frac{\delta}{\delta \dot{A}^i_{0}} \right],
\label{gaupal}
\end{eqnarray}
where in $Z_7$ and $Z_8$ we have chosen the parameters to be $-\rho$ and $-\tau$, respectively. Notice that the vector (\ref{gaupal}) kills the action (\ref{palati}) up to boundary terms. The arbitrariness  of the components of this vector involving $\varepsilon^i_{30}$ and $\varepsilon^i_{40}$ just reflects the fact that ${\dot e}^i_0$ and ${\dot A}^i_0$ are not present in the corresponding Lagrangian [see Eq. (\ref{palati2+1})]. Now, in order to fix  $\varepsilon^i_{30}$ and $\varepsilon^i_{40}$, we demand the vector field to be of the form
\begin{eqnarray}
\tilde{X}&=&\int \d^2x \left[ \delta_{\varepsilon} e^i_{\mu}\frac{\delta}{\delta e^i_{\mu}}+\delta_{\varepsilon} A^i_{\mu}\frac{\delta}{\delta A^i_{\mu}}+\partial_0(\delta_{\varepsilon} e^i_{\mu}) \frac{\delta}{\delta \dot{e}^i_{\mu}} + \partial_0 \left( \delta_{\varepsilon} A^i_{\mu}\right)\frac{\delta}{\delta \dot{A}^i_{\mu}} \right],\label{mia0}
\end{eqnarray} 
which corresponds to the second step described in Subsection \ref{GT}. This yields the following relations between the parameters
\begin{eqnarray} 
 \varepsilon^i_{30}&=& \partial_0 \left(\mathcal{D}_0 \rho^i+f^i{}_{jk} e^j_0\tau^k\right),\nonumber\\
 \varepsilon^i_{40} &=& \partial_0\left( \mathcal{D}_0 \tau^i- \Lambda \epsilon^i{}_{jk} e^j_0\rho^k \right).\label{rel1}
\end{eqnarray} 
Consequently, the gauge vector field becomes
\begin{eqnarray}
\tilde{X}&=&\int \d^2x  \left[ \left( \mathcal{D}_{\mu} \rho^i+f^i{}_{jk} e^j_{\mu} \tau^k \right)\frac{\delta}{\delta e^i_{\mu}} + \left(\mathcal{D}_{\mu}\tau^i - \Lambda \epsilon^i{}_{jk} e^j_{\mu}\rho^k \right) \frac{\delta}{\delta A^i_{\mu}} \right.\nonumber\\
&&  + \partial_0 \left( \mathcal{D}_{\mu} \rho^i+f^i{}_{jk} e^j_{\mu} \tau^k \right) \frac{\delta}{\delta \dot{e}^i_{\mu}}   \left.+ \partial_0 \left( \mathcal{D}_{\mu}\tau^i - \Lambda \epsilon^i{}_{jk} e^j_{\mu}\rho^k \right) \frac{\delta}{\delta \dot{A}^i_{\mu}} \right].
\end{eqnarray}
Therefore, the off-shell gauge transformations read $\delta_{\varepsilon}e^i_{\mu}=\mathcal{D}_{\mu} \rho^i + f^i{}_{jk} e^j_{\mu}\tau^k$ and $
\delta_{\varepsilon}A^i_{\mu}= \mathcal{D}_{\mu}\tau^i- \Lambda \epsilon^i{}_{jk} e^j_{\mu}\rho^k$, which in terms of differential forms can be written as 

\begin{eqnarray}
\delta_{\varepsilon}e^i&=&\mathcal{D} \rho^i + f^i{}_{jk} e^j\tau^k, \nonumber\\
\delta_{\varepsilon}A^i&=& \mathcal{D}\tau^i- \Lambda \epsilon^i{}_{jk} e^j\rho^k.\label{gatrpala}
\end{eqnarray}
The Noether identities (from which these gauge transformations can also be read off) are given by $\mathcal{D} {\mathcal E}^i -\Lambda  \epsilon^i\,_{jk} e^j \wedge {\mathcal E}'^k  =0$ and  $\mathcal{D} {\mathcal E}'^i+ f^i\,_{jk}  e^j \wedge {\mathcal E}^k =0$, respectively.\cite{Celada} Moreover, the action (\ref{palati}) is also manifestly invariant under spacetime diffeomorphisms. Nevertheless, diffeomorphism invariance of the action is not independent from (\ref{gatrpala}). Indeed, using that the Lie derivative for any differential form $\sigma$ along the vector field $\xi$ satisfies Cartan's formula
\begin{eqnarray}
\pounds_{\xi} \sigma&=& \xi \cdot\d \sigma + \d \left( \xi \cdot \sigma \right), \label{Cartan}
\end{eqnarray}
where the dot `$\cdot$' stands for the contraction of the vector $\xi$ and a differential form, we have that
\begin{eqnarray}
\pounds_{\xi} e^i&=&\mathcal{D} \left( \xi \cdot e^i \right) + f^i{}_{jk} e^j_{\mu}\left( \xi \cdot A^k \right)+ \xi \cdot \left( \mathcal{D} e^i\right), \nonumber\\
 \pounds_{\xi} A^i &=&\mathcal{D}\left( \xi\cdot A^i \right)- \Lambda \epsilon^i{}_{jk} e^j \left(  \xi \cdot e^k \right) +\xi \cdot \left( F^i- \frac{\Lambda}{2} \epsilon^i{}_{jk} e^j\wedge e^k \right), \label{diff3d}
\end{eqnarray}
and so an infinitesimal diffeomorphism is a linear combination of the gauge transformation \eqref{gatrpala} with field-dependent parameters $\rho^i=\xi \cdot e^i $ and $\tau^i= \xi \cdot A^i$ plus terms that are proportional to the variational derivatives \eqref{imp2}. Therefore, Eqs. \eqref{diff3d} say that the diffeomorphism transformation corresponds to the transformations \eqref{gatrpala} plus a trivial one (gauge transformations that vanish on-shell are called trivial gauge transformations\cite{Teitelboim}).

On the other hand, from (\ref{gatrpala}) the gauge transformations involve $\rho^i$ and $\tau^i$, which are the independent gauge parameters. Therefore, $g=2 \times{\rm dim} \, \mathfrak{g}$. Also, from  (\ref{gatrpala}) we have that they also involve their first time derivative, then we have $e=4 \times{\rm dim} \, \mathfrak{g}$. Furthermore, the map (\ref{map}) allows us to know the number of first- and second-class constraints 
\begin{eqnarray}
N_1 &=& e = 4 \times {\rm dim}\, \mathfrak{g},\nonumber\\
N_2 &=& l + g - e = (6+2-4) \times {\rm dim}\, \mathfrak{g} \nonumber \\
&=& 4 \times {\rm dim}\, \mathfrak{g}, \label{mapita}
\end{eqnarray}
which must appear in its Dirac's canonical analysis.

\subsubsection{Three-dimensional generalized Palatini theory with $\Lambda =0$}

Consider now the case without cosmological constant. The analysis can be obtained by setting $\Lambda=0$ in Subsections \ref{IIIAI} and \ref{IIIAII}. Then, we only quote the results. The points of the tangent bundle $T\cal{C}$ are locally labeled with the same variables $(e^i_{\mu}, A^i_{\mu}, {\dot e}^i_{\mu}, {\dot A}^i_{\mu})$, and the symplectic structure in this case is the same as in (\ref{psp}). Therefore, it has the same null vectors (\ref{inv}). As a consequence of the fact that the energy terms containing $\Lambda$ in \eqref{psp} do not appear anymore, the projectable constraints \eqref{etiqueta} reduce to
\begin{eqnarray}
\varphi_{1i}:= \epsilon^{0ab} F_{iab}\approx 0, \qquad \varphi_{2i}:=\epsilon^{0ab}\mathcal{D}_a e_{ib} \approx 0.
\end{eqnarray}
The vectors associated with these constraints that are tangent to the Lagrangian constraint surface, i.e., the null associated vector fields, are 
\begin{eqnarray}
Z_7[N^i]&=& -\int \d^2x \left[ \left( \mathcal{D}_{\mu} N^i \right)\frac{\delta}{\delta e^i_{\mu}} + \partial_0 \left( \mathcal{D}_{a} N^i \right) \frac{\delta}{\delta \dot{e}^i_{a}} \right], \nonumber\\
Z_8 [M^i]&=& -\int \d^2x \left[ f^i{}_{jk} e^j_{\mu} M^k \frac{\delta}{\delta e^i_{\mu}}+ \mathcal{D}_{\mu} M^i \frac{\delta}{\delta A^i_{\mu}} + \partial_0\left( f^i{}_{jk} e^j_{a} M^k \right) \frac{\delta}{\delta \dot{e}^i_{a}}\right.\nonumber\\
&&\left. + \partial_0\left( \mathcal{D}_{a} M^i \right) \frac{\delta}{\delta \dot{A}^i_{a}} \right],
\end{eqnarray}
respectively. Also, the non-projectable constraints  \eqref{cnp3d}  are now
\begin{eqnarray}
\varphi_{1}^{ia} := \epsilon^{a\mu \nu} F^i_{\mu \nu}\approx0, \qquad
\varphi_{2}^{ia}  := \epsilon^{a \mu \nu} \mathcal{D}_{\mu} e^i_{\nu}\approx0.
\end{eqnarray}
{\it Degree of freedom count}. Notice that the number of the total null vectors and Lagrangian constraints do not change. Therefore, the counting of the physical degrees of freedom is the same as in the case for $\Lambda \neq 0$. Furthermore, the gauge transformations now read (the corresponding Noether identities are $\mathcal{D} {\mathcal E}^i=0$ and  $\mathcal{D} {\mathcal E}'^i+ f^i\,_{jk}  e^j  \wedge {\mathcal E}^k =0$\cite{Celada})
\begin{equation}
\delta_{\varepsilon}e^i=\mathcal{D} \rho^i + f^i{}_{jk} e^j \tau^k, \qquad \delta_{\varepsilon}A^i= \mathcal{D}\tau^i.
\end{equation}
The discussion of the diffeomorphism symmetry is along the same lines as in Subsection \ref{IIIAII} just by setting $\Lambda=0$ there. 

Finally, it follows that the map (\ref{mapita}) also holds for $\Lambda =0$, i.e., the number of first- and second-class constraints that appear in Dirac's canonical analysis of the generalized Palatini theory (or simply BF theory) is the same for  $\Lambda =0$ and $\Lambda \neq 0$.

%\\\\\\\\\\\\\\\\\\\\\\\\\\\\\\\\\\
\subsection{Witten's exotic action}

Now, we present Witten's exotic action\cite{Witten} but based on $G$, so we are considering a generalized action. As we will see, this theory is classically equivalent to the generalized Palatini theory, in the sense that both actions yield the same equations of motion if we choose $\epsilon^{ijk}= f^i{}_{lm}k^{lj}k^{mk}$ in (\ref{imp2}). The action is given by 
\begin{eqnarray}
S[e, A] &=& \int_{M} \left( A_i \wedge \d A^i + \frac{1}{3} f^{ijk} A_i\wedge A_j\wedge A_k -\Lambda \, e_i \wedge \mathcal{D} e^i \right), \label{witten} 
\end{eqnarray}
where $e^i=e^i_{\mu} \d x^{\mu}$ is a $\mathfrak{g}$-valued $1$-form, $A^i=A^i_{\mu} \d x^{\mu}$ is a $\mathfrak{g}$-valued connection $1$-form, and 
$\mathcal{D} v^i:= {\rm d} v^i+f^i{}_{jk} A^j \wedge v^k$ (as in Subsection \ref{ejem1}). Notice that this action and the generalized Palatini action (\ref{palati}) are functionals of the same field variables. Thus, we have $N= 6 \times{\rm dim} \, \mathfrak{g}$ configuration variables. The variation of the action principle (\ref{witten}) with respect to the independent variables yields the variational derivatives ${\mathcal E}'^i$ and ${\mathcal E}^i$ given by
\begin{eqnarray}
\delta e^i &:&\quad {\mathcal E}'^i:= -2\Lambda \mathcal{D}e^i, %\Leftrightarrow \epsilon^{\mu\nu\rho} \mathcal{D}_{\nu}e^i_{\rho} =0, 
\nonumber\\
\delta A^i &:& \quad {\mathcal E}^i := 2F^i-\Lambda f^i\,_{jk} e^j \wedge e^k.
%\nonumber\\ &&\Leftrightarrow\epsilon^{\mu\nu\rho} \left( F^i_{\nu\rho}-\Lambda f^{ijk}e_{j \nu} e_{k \rho} \right)=0.
 \label{witimp2}
\end{eqnarray}
where $F^i = \d A^i + \frac{1}{2} f^i\,_{jk} A^j \wedge A^k = \frac{1}{2} F^i_{\mu \nu} \d x^{\mu}\wedge \d x^{\nu}$ is the curvature of $A^i$. Then, the equations of motion are
\begin{eqnarray}
{\mathcal E}'^i =0, \qquad {\mathcal E}^{i}=0. \label{emwa}
\end{eqnarray}
These equations of motion, up to global constant factors, coincide with those given in Eq. (\ref{empa}) if we take $\epsilon^{ijk}= f^i{}_{lm}k^{lj}k^{mk}$ there. Hence, under this choice, both actions describe the same theory at classical level. Notice that we could have defined a similar action to \eqref{witten} but with some tensor $\epsilon^{ijk}$ (instead of $f^{ijk}$) and that such an action would not be equivalent in general to Palatini's action \eqref{palati}. Here, we follow the standard approach to Witten's action.

By performing the $2+1$ decomposition of the fields, the action (\ref{witten}) acquires the form
\begin{eqnarray}
S[e, A]&=&\int_{M} \d^3x \, \epsilon^{0ab} \left[\Lambda e_{ia}\dot{e}^i_b-A_{ia} \dot{A}^i_b-2\Lambda e_{i0}\mathcal{D}_a e^i_b %\right. \nonumber\\&&\left.
 + A_{i0}\left( F^i_{ab}-\Lambda f^{ijk}e_{ja}e_{kb}\right) \right].
\end{eqnarray} 
{\it Lagrangian constraints}. The points of the tangent bundle $T {\mathcal C}$ are locally labeled with $12 \times{\rm dim} \, \mathfrak{g}$ variables given by $(e^i_{\mu}, A^i_{\mu}, {\dot e}^i_{\mu}, {\dot A}^i_{\mu})$, as in the preceding example. The corresponding symplectic structure (\ref{gp1}), energy (\ref{energy}), and the differential of the energy are
\begin{eqnarray}\label{witpsp}
\Omega &=&\int \d^{2}x \, \epsilon^{0ab} \left(\Lambda\, \ed e_{ia} \ew \ed e^i_b- \ed A_{ia} \ew \ed A^i_b \right), \nonumber\\
E &=&\int \d^{2}x \, \epsilon^{0ab} \left[-A_{i0}\left( F^i_{ab}-\Lambda f^{ijk}e_{ja}e_{kb}\right) 
+2\Lambda e_{i0} \mathcal{D}_a e^i_{b} \right], \nonumber \\
\ed E&=& \int \d^{2}x \, \epsilon^{0ab}\left[ - \left( F^i_{ab}-\Lambda f^{ijk}e_{ja}e_{kb}\right) \ed A_{i0} + 2\left(\mathcal{D}_aA^i_0+ \Lambda f^{ijk} e_{j0}e_{ka}\right)\ed A_{ib}  \right. \nonumber\\
&& \left.+2\Lambda\mathcal{D}_ae^i_{b}\, \ed e_{i0}  -2\Lambda\left( \mathcal{D}_a e^i_{0} -f^i{}_{jk} A^j_0e^k_{a} \right)\ed e_{ib} \right].
\end{eqnarray} 
Continuing with the formalism, we look for the vector field $X$ in (\ref{gp2+})
\begin{equation}
X:=\int \d^{2}x \left( \alpha^i_ {\mu} \frac{\delta}{\delta e^i_{\mu} }+\alpha'^i_{\mu} \frac{\delta}{\delta A^i_{\mu} } +\beta^i_{\mu} \frac{\delta}{\delta\dot{e}^i_{\mu} }+\beta'^i_{\mu} \frac{\delta}{\delta \dot{A}^i_{\mu} }\right),
\end{equation}
and we obtain
\begin{eqnarray}
\epsilon^{0ab}\alpha^i_{a} &=& \epsilon^{0ab}\left( \mathcal{D}_a e^i_{0}-f^i{}_{jk} A^j_0e^k_{a} \right), \nonumber\\
\epsilon^{0ab} \alpha'^i_{a}&=& \epsilon^{0ab} \left( \mathcal{D}_a A^i_0+ \Lambda f^{ijk} e_{j0}e_{ka}\right), \nonumber\\
 0&=& \epsilon^{0ab} \left( F^i_{ab}-\Lambda f^{ijk}e_{ja}e_{kb}\right),\nonumber \\
 0&=&-2\Lambda\epsilon^{0ab} \mathcal{D}_ae^i_{b},\label{witeapalg}
\end{eqnarray}
which agree with (\ref{witimp2}) once we substitute $\alpha_{ia}= \dot{e}_{ia}$ and $\alpha'_{ia}= \dot{A}_{ia}$ in the LHS of the first two
lines of (\ref{witeapalg}). In spite of the fact that the symplectic structure (\ref{witpsp}) is different from the symplectic structure (\ref{psp}) of the Palatini theory, they have the same rank, $4 \times{\rm dim} \, \mathfrak{g}$, and share the same basis of $8 \times{\rm dim} \, \mathfrak{g}$ null vectors given by (\ref{inv}). Furthermore, among them, only $ Z_{1a}$ and $Z_{2a}$ generate $2 \times{\rm dim} \, \mathfrak{g}$ {\it projectable} Lagrangian constraints, given by
\begin{eqnarray}
\int \d^2x \, \varphi_{1i} & := & Z_{1i}\cdot \ed E= \int \d^2x \, 2\Lambda \epsilon^{0ab} \mathcal{D}_ae_{ib} \approx 0, \nonumber\\
\int \d^2x \, \varphi_{2i} & :=& Z_{2i}\cdot \ed E = -\int \d^2x \, \epsilon^{0ab} \left( F_{iab}-\Lambda f_{ijk}e^j_{a}e^k_{b}\right) \approx 0.
\end{eqnarray}
These constraints are, up to global constant factors, those found for the generalized Palatini theory \eqref{etiqueta} (if we therein use $\epsilon^{ijk}= f^i{}_{lm}k^{lj}k^{mk}$). Notice that the constraint generated by $Z_{1i}$ corresponds to the constraint generated by $Z_{2i}$ in the generalized Palatini theory (and viceversa). As in Subsection \ref{ejem1}, for this theory, there are no more projectable Lagrangian constraints. 

Now, we search for the non-projectable Lagrangian constraints. The requirement $\alpha_{ia}= \dot{e}_{ia}$ and $\alpha'_{ia}= \dot{A}_{ia}$ in (\ref{witeapalg}) implies the new $ 4 \times{\rm dim} \, \mathfrak{g}$ {\it non-projectable} Lagrangian constraints,
\begin{eqnarray}
\varphi_{1}^{ia} & := & \epsilon^{a\mu \nu} \left( F^i_{\mu \nu}-\Lambda f^{i}{}_{jk} e^j_{\mu} e^k_{\nu} \right)\approx 0, \nonumber\\
\varphi_{2}^{ia} & :=& \epsilon^{a \mu \nu} \mathcal{D}_{\mu} e^i_{\nu}\approx 0.
\end{eqnarray}
Notice that these are the same non-projectable constraints \eqref{cnp3d} of the generalized Palatini theory if we use $\epsilon^{ijk}= f^i{}_{lm}k^{lj}k^{mk}$ there. Therefore, we have $l= (2+4) \times{\rm dim} \, \mathfrak{g}=6 \times{\rm dim} \, \mathfrak{g}$ Lagrangian constraints.

\subsubsection{Degree of freedom count} %Witten

{\it Associated vector fields}. Using the test fields $N^i$ and $M^i$, we smear the projectable constraints,
\begin{equation}
\varphi_1[N^i]:=\int \d^{2}x \, N^i \varphi_{1i}, \qquad \varphi_2[M^i]:=\int \d^{2}x  \, M^i \varphi_{2i}. 
\end{equation}
Therefore, the associated vectors to these constraints, modulo the null vectors \eqref{inv}, are
\begin{eqnarray}\label{avw}
Z_5[N^i]&=& -\int \d^{2}x \left( \mathcal{D}_a N^i \frac{\delta}{\delta e^i_a}-\Lambda f^i{}_{jk} e^j_a N^k \frac{\delta}{\delta A^i_a} \right), \nonumber\\
Z_6 [M^i]&=&- \int \d^{2}x\left( f^i{}_{jk} e^j_a M^k \frac{\delta}{\delta e^i_a}+ \mathcal{D}_a M^i \frac{\delta}{\delta A^i_a} \right).\label{witinv}
\end{eqnarray}
Notice that these associated vectors coincide with (\ref{nullvecpa}) of the Palatini theory if we choose $\epsilon^{ijk}= f^i{}_{lm}k^{lj}k^{mk}$ there. Notice that the constraint associated with $Z_5 [N^i]$ corresponds to the constraint associated with $Z_6 [M^i]$ in the generalized Palatini theory (and viceversa) because they satisfy $Z_5[N^i] \cdot \Omega = \,\ed \varphi_1[N^i] $ and $Z_6 [M^i] \cdot \Omega=\, \ed \varphi_2[M^i]$ (modulo a surface term).

{\it Null associated vector fields}. Now we check that they generate null associated vector fields, i.e., that there are linear combinations of them and the original null vectors (\ref{inv}) that are tangent to the Lagrangian constraint surface. Nevertheless, as the reader can realize, this calculation is already done because all the information (constraints, null vectors, and associated vectors) found for this theory coincides with that found in the analysis of Palatini's action (with $\epsilon^{ijk}= f^i{}_{lm}k^{lj}k^{mk}$). Therefore, the null associated vectors are
\begin{eqnarray}\label{Z7Z8}
Z_7[N^i]&=& -\int \d^2x \left[ \left( \mathcal{D}_{\mu} N^i \right)\frac{\delta}{\delta e^i_{\mu}} + \left( - \Lambda f^i{}_{jk} e^j_{\mu} N^k \right) \frac{\delta}{\delta A^i_{\mu}} + \partial_0 \left( \mathcal{D}_{a} N^i \right) \frac{\delta}{\delta \dot{e}^i_{a}} \right.\nonumber\\
&& \left.   + \partial_0 \left(- \Lambda f^i{}_{jk} e^j_a N^k \right) \frac{\delta}{\delta \dot{A}^i_{a}} \right], \nonumber\\
Z_8 [M^i]&=& -\int \d^2x \left[ f^i{}_{jk} e^j_{\mu} M^k \frac{\delta}{\delta e^i_{\mu}}+ \mathcal{D}_{\mu} M^i \frac{\delta}{\delta A^i_{\mu}} + \partial_0\left( f^i{}_{jk} e^j_{a} M^k \right) \frac{\delta}{\delta \dot{e}^i_{a}} \right.\nonumber\\
&& \left.  + \partial_0\left( \mathcal{D}_{a} M^i \right) \frac{\delta}{\delta \dot{A}^i_{a}} \right].
\end{eqnarray}

{\it Degree of freedom count}. Now, we can make the counting of the physical degrees of freedom: the number of field variables $e^i_{\mu}$ and $A^i_{\mu}$ is $N=6 \times{\rm dim} \, \mathfrak{g}$, the total number of null vectors, given by (\ref{inv}) and (\ref{Z7Z8}), is $B=(8+2) \times{\rm dim} \, \mathfrak{g}= 10 \times{\rm dim} \, \mathfrak{g}$, and the number of linearly independent combinations of constraints [to get the associated vector fields \eqref{avw}] is $C=2 \times{\rm dim} \, \mathfrak{g}$. Therefore, using \eqref{pdfc}, the number of physical degrees of freedom is
\begin{equation}
\frac{1}{2} {\rm Rank} \, \iota^*\Omega = \left[6-\frac{1}{2} \left(10+2 \right) \right]\times {\rm dim} \, \mathfrak{g}=0.
\end{equation}
Furthermore, by substituting this result into the right-hand side of Eq. (\ref{gl1}) we get $g+e=2N - l = (2 (6)- 6) \times{\rm dim} \, \mathfrak{g} = 6 \times{\rm dim} \, \mathfrak{g}$, which is the number of independent null vectors of $\iota^* \Omega$.

\subsubsection{Off-shell gauge transformations}

In order to find the gauge transformations we must form a vector field with the null associated vector fields \eqref{Z7Z8} and  the null vector fields, $Z_{3i}^{0}$ and $Z_{4i}^{0}$, which are tangent to the Lagrangian constraint surface. However, we have seen that these vectors coincide with those found in the analysis of the Palatini action with $\epsilon^{ijk}= f^i{}_{lm}k^{lj}k^{mk}$. Then, the analysis follows the same steps, and then the vector field $\tilde{X}$ is given by
\begin{eqnarray}
\tilde{X}&=&\int \d^2x  \left[ \left( \mathcal{D}_{\mu} \rho^i+f^i{}_{jk} e^j_{\mu} \tau^k \right)\frac{\delta}{\delta e^i_{\mu}} + \left(\mathcal{D}_{\mu}\tau^i - \Lambda f^i{}_{jk} e^j_{\mu}\rho^k \right) \frac{\delta}{\delta A^i_{\mu}} \right.\nonumber\\
&&   + \partial_0 \left( \mathcal{D}_{\mu} \rho^i+f^i{}_{jk} e^j_{\mu} \tau^k \right) \frac{\delta}{\delta \dot{e}^i_{\mu}}  \left.+ \partial_0 \left( \mathcal{D}_{\mu}\tau^i - \Lambda f^i{}_{jk} e^j_{\mu}\rho^k \right) \frac{\delta}{\delta \dot{A}^i_{\mu}} \right]. \label{witgvr}
\end{eqnarray}
Therefore, the gauge transformations in terms of differential forms read
\begin{eqnarray}
\delta_{\varepsilon}e^i &=& \mathcal{D} \rho^i + f^i{}_{jk} e^j \tau^k, \nonumber\\
\delta_{\varepsilon}A^i &=& \mathcal{D} \tau^i- \Lambda f^i{}_{jk} e^j\rho^k.
\end{eqnarray}
These gauge transformations coincide with (\ref{gatrpala}) (with $\epsilon^{ijk}= f^i{}_{lm}k^{lj}k^{mk}$ there). As a result, under this identification, Palatini and Witten's exotic actions also share the same gauge transformations. The Noether identities (from which the gauge transformations can be read off) are given by $\mathcal{D} {\mathcal E}'^i -\Lambda  f^i\,_{jk} e^j \wedge {\mathcal E}^k  =0, $ and  $\mathcal{D} {\mathcal E}^i+ f^i\,_{jk} e^j \wedge {\mathcal E}'^k =0$ (cf. with the case of Palatini's action). Diffeomorphism invariance of Witten's exotic action (\ref{witten}) can also be analyzed along the same lines that the case of the generalized Palatini action discussed in Subsection \ref{ejem1}. Furthermore, the fact that both action principles (\ref{palati}) and (\ref{witten}) yield the same equations of motion and share the gauge transformations allows us to consider a linear combination of both action principles to get a new action principle, which has been analyzed in Ref.~\onlinecite{Livine} and has been related to the presence of an Immirzi-like parameter in three dimensions.

Finally, as a consequence that both actions share the gauge transformations and the number of Lagrangian constraints, their Lagrangian parameters coincide. Therefore, using the map \eqref{map} we have that the generalized Witten action involves
\begin{eqnarray}
N_1 &=& e = 4 \times {\rm dim}\, \mathfrak{g},\nonumber\\
N_2 &=& l + g - e = (6+2-4) \times {\rm dim}\, \mathfrak{g} \nonumber \\
&=& 4 \times {\rm dim}\, \mathfrak{g},
\end{eqnarray}
first- and second-class constraints, respectively, if its Dirac's canonical analysis is performed. 

On the other hand, we have shown that generalized Palatini and Witten's actions, given by (\ref{palati}) and (\ref{witten}), yield the same equations of motion for $\Lambda \neq 0$ and the identification $\epsilon^{ijk}= f^i{}_{lm}k^{lj}k^{mk}$, but this is not true if we set the cosmological constant equal to zero in both action principles. In fact, by setting $\Lambda=0$ in the generalized Palatini action (\ref{palati}) we get an action that depends on the same field variables $e^i$ and $A^i$ and still describes the same theory just without cosmological constant. On the other hand, if we set $\Lambda=0$ in generalized Witten's action (\ref{witten}), we obtain a theory that only depends on the connection $A^i$: Chern-Simons theory \cite{Chern,Naka, Bla} (for the relationship between the Palatini and Chern-Simons actions, see Refs.~\onlinecite{Romano, Witten}). Chern-Simons theory is relevant in different contexts and we analyze it in Subsection \ref{ejem3}.

%\\\\\\\\\\\\\\\\\\\\\\\\\\\\\\\\
\subsection{Chern-Simons theory}\label{ejem3}
In this subsection, we analyze Chern-Simons theory, which is defined by the action principle
\begin{eqnarray}
S[A] = \int_{M}  A_i \wedge \left(  \d A^i + \frac{1}{3} f^{ijk}A_j\wedge A_k \right), \label{chern}
%S[A] = \int_{M} d^3x \left( A_i \wedge \d A^i + \frac{1}{3} f^{ijk} A_i\wedge A_j\wedge A_k \right), \label{chern}
\end{eqnarray}
which is a functional of $N= 3 \times{\rm dim} \, \mathfrak{g}$ configuration variables contained in the $\mathfrak{g}$-valued  connection 1-form $A^i$. Its variation with respect to $A^i$ yields the variational derivative ${\mathcal E}^i$ 
\begin{eqnarray}
\delta A^i &:& {\mathcal E}^i:= 2F^i, \label{cherimp}
\end{eqnarray}
where $F^i$ is the curvature $F^i = \d A^i + \frac12 f^i\,_{jk} A^j \wedge A^k$. So, the equations of motion are ${\mathcal E}^i=0$. 

We begin the analysis by making the spacetime decomposition of the connection so that the action (\ref{cherimp}) acquires the form
\begin{equation}
S[A]=\int_{M} \d^3x \, \epsilon^{0ab}\left( -A_{ia} \dot{A}^i_b +A_{i0} F^i_{ab} \right).
\end{equation}
{\it Lagrangian constraints}. In this case, the points of the tangent bundle $T\cal{C}$ are locally labeled with
$6 \times{\rm dim} \, \mathfrak{g}$ variables given by $(A^i_\mu, \dot{A}^i_\mu)$. The corresponding symplectic structure (\ref{gp1}), energy (\ref{energy}), and the differential of the energy are
\begin{eqnarray}\label{chepsp}
\Omega &=&-\int \d^{2}x \, \epsilon^{0ab} \ed A_{ia} \ew \ed A^i_b, \nonumber\\
E &=&-\int \d^{2}x \, \epsilon^{0ab}A_{i0} F^i_{ab}, \nonumber \\
\ed E&=& \int \d^{2}x \, \epsilon^{0ab} \left( - F^i_{ab} \, \ed A_{i0} + 2\mathcal{D}_a A^i_0 \,\ed A_{ib} \right),
\end{eqnarray} 
where $\mathcal{D}_a$ are the components of the covariant derivative associated with $A^i$. We look for the vector field $X:=\int \d^{2}x \left( \alpha^i_ {\mu} \frac{\delta}{\delta A^i_{\mu} }+\beta^i_{\mu} \frac{\delta}{\delta\dot{A}^i_{\mu} }\right)$ in (\ref{gp2+}), and we get 
\begin{eqnarray}
\epsilon^{0ab} \alpha^i_{a}&=& \epsilon^{0ab} \mathcal{D}_a A^i_0, \nonumber\\
 0&=& \epsilon^{0ab} F^i_{ab},\label{chernapalg}
\end{eqnarray}
which of course coincide with (\ref{cherimp}), once we substitute $\alpha_{ia}= \dot{A}_{ia}$ in the LHS of the first one. We will come back to these equations below when we search for the non-projectable constraints. On the other hand, a basis of $\ker \Omega$ is given by the $4 \times{\rm dim} \, \mathfrak{g}$  vector fields,
\begin{eqnarray}\label{cherinv}
Z_{i} := \int \d^{2}x \, \frac{\delta}{\delta A^i_0}, \qquad Z_{i}^{\mu}:= \int \d^{2}x \, \frac{\delta}{\delta \dot{A}^i_{\mu}}.
\end{eqnarray} 
However, only $Z_{i}$ generates  $1 \times{\rm dim} \, \mathfrak{g}$ projectable Lagrangian constraints given by
\begin{eqnarray}
\int \d^2x \, \varphi_{i} & :=& Z_{i}\cdot \ed E= -\int \d^2x \, \epsilon^{0ab} F_{iab} \approx 0.
\end{eqnarray}
For this theory, there are no more projectable Lagrangian constraints. 

Now, we search for the non-projectable Lagrangian constraints. The requirement $\alpha_{ia}= \dot{A}_{ia}$ in (\ref{chernapalg}) implies that the fields must satisfy the $2 \times{\rm dim} \, \mathfrak{g}$ non-projectable Lagrangian constraint
\begin{eqnarray}\label{npccs}
\varphi^{ia} & := & \epsilon^{a\mu \nu}F^i_{\mu \nu}\approx 0.
\end{eqnarray}
Thus, this theory has $l= (1+2) \times{\rm dim} \, \mathfrak{g} =  3 \times{\rm dim} \, \mathfrak{g}$ Lagrangian constraints.

\subsubsection{Degree of freedom count} %Chern

{\it Associated vector fields}. We use the test fields $N^i$ to define the smeared projectable constraints,
%Notice that they are part of the equations of motion,
\begin{equation} \label{spccs}
\varphi[N^i]:=\int \d^{2}x \, N^i \varphi_{i}.
\end{equation}
Its associated vectors, modulo the null vectors \eqref{cherinv}, are
\begin{eqnarray}
Z_1[N^i]&=& \int \d^{2}x \, \mathcal{D}_a N^i \frac{\delta}{\delta A^i_a}\label{chetinv}
\end{eqnarray}
because, they satisfy $Z_1[N^i] \cdot \Omega = -\,\ed \varphi[N^i] $ modulo a surface term.

{\it Null associated vector fields}. The associated vector fields (\ref{chetinv}) are associated
to Lagrangian constraints that are part of the equations of motion (\ref{cherimp}); therefore, we
expect that they are indeed null associated vector fields. Therefore, if we used that criterion, we could
make the counting of the physical degrees of freedom right now. Continuing with the general procedure,
we form  with $Z_1$ and the original null vectors \eqref{cherinv} the vector field
\begin{equation}
Z_2:=Z_1[N^i]+ Z_{i}[M^i]+ Z_{i}^{\mu}[ M^i_{\mu}], \label{nvfcs}
\end{equation}
where $M^i$ and $M^i_{\mu}$ are smearing functions. Now, we demand the vector fields (\ref{nvfcs}) to be tangent to the Lagrangian constraint surface, i.e., $Z_2$ have to override all the Lagrangian constraints \eqref{spccs} and the smeared version of \eqref{npccs}. This requirement yields 
\begin{equation}
M^{i}\approx -\mathcal{D}_0 N^i , \qquad M^{i}_{a}\approx -\partial_0 \left( \mathcal{D}_a N^i\right), 
\end{equation}
and the parameter $M^{i}_{0}$ is left arbitrary ($M^{i}_{0}$ is not fixed because it smears a vector field $ Z_{i}^{0}$, which is tangent to the Lagrangian constraint surface). Thus, we have $1 \times{\rm dim} \, \mathfrak{g}$ null associated vectors given by
\begin{equation}
Z_2 [N^i]= -\int \d^2x \left[ \mathcal{D}_{\mu} N^i \frac{\delta}{\delta A^i_{\mu}} + \partial_0\left( \mathcal{D}_{a} N^i \right) \frac{\delta}{\delta \dot{A}^i_{a}} \right]. \label{gtcs}
\end{equation}

{\it Degree of freedom count}. We can now make the counting of the physical degrees of freedom. In summary, the number of field variables $A^i_\mu$ is $N=3 \times{\rm dim} \, \mathfrak{g}$, we have $B=5 \times{\rm dim} \, \mathfrak{g}$ null vectors [$4 \times{\rm dim} \, \mathfrak{g}$ original ones \eqref{cherinv} and $1 \times{\rm dim} \, \mathfrak{g}$ null associated vectors \eqref{gtcs}], and we have $C=1 \times{\rm dim} \, \mathfrak{g}$ associated vector fields \eqref{chetinv}. Therefore, the counting of the physical degrees of freedom given by \eqref{pdfc} is 
\begin{equation}
\frac{1}{2} {\rm Rank} \, \iota^*\Omega = \left [ 3-\frac{1}{2} \left(5+1 \right) \right ] \times {\rm dim} \, \mathfrak{g}=0.
\end{equation}
%Then, as we have anticipated, there are no degrees of freedom per point of spacetime.

As a consequence of Eq. (\ref{gl1}), we have that $g+e=3 \times{\rm dim} \, \mathfrak{g}$.

\subsubsection{Off-shell gauge transformations}
The reader must follow the two steps explained in Subsection \ref{GT} to obtain the gauge transformations. First, we use only the original null vector fields $Z_{i}^{0}$ that are tangent to the Lagrangian constraint surface and the null associated vectors (\ref{gtcs}) to form the vector field, 
\begin{eqnarray}
\tilde{X}&:=& Z^0_{i}[M^i_0]+Z_2[-\varepsilon^i] \nonumber\\
&:=&\int \d^{2}x \left[ M^i_0\frac{\delta}{\delta \dot{A}^i_{0}} + 
\mathcal{D}_{\mu} \varepsilon^i \frac{\delta}{\delta A^i_{\mu}} + \partial_0\left( \mathcal{D}_{a} \varepsilon^i \right) \frac{\delta}{\delta \dot{A}^i_{a}} \right],
\end{eqnarray} where we have used $-\varepsilon^i$ as smearing functions, in order to agree with the standard notation. Continuing with the procedure, in the second step of Sec. \ref{GT}, we demand the vector field $\tilde{X}$ to be of the form
\begin{eqnarray}
\tilde{X}&=&\int \d^2x \left[\delta_{\varepsilon} A^i_{\mu}\frac{\delta}{\delta A^i_{\mu}}+ \partial_0 \left( \delta_{\varepsilon} A^i_{\mu}\right)\frac{\delta}{\delta \dot{A}^i_{\mu}} \right],\label{mial01}
\end{eqnarray}
which yields $M^i_{0} = \partial_0 \left(\mathcal{D}_{0}\varepsilon^i\right)$. Therefore, the gauge transformations read 
\begin{eqnarray}
\delta_{\varepsilon} A^i &=& \mathcal{D} \varepsilon^i.\label{gauchern}
\end{eqnarray}
The Noether identity from which this gauge transformation can be read off is $\mathcal{D} {\mathcal E}^i=0$. Besides, the action (\ref{chern}) is also invariant under spacetime diffeomorphisms. As in the previous examples, diffeomorphism invariance is not an independent gauge transformation. In fact, using Cartan's formula \eqref{Cartan}, we have that
\begin{eqnarray}
\pounds_{\xi} A^i &=& \mathcal{D}\left( \xi\cdot A^i \right) +\xi \cdot F^i. \label{diffchern}
\end{eqnarray}
Therefore, an infinitesimal diffeomorphism can be identified with the gauge transformations \eqref{gauchern} with the (field-dependent) parameters $\varepsilon^i=\xi \cdot A^i$ plus a trivial transformation.

On the other hand, notice that the gauge transformations \eqref{gauchern} involve the parameters $\varepsilon^i$, then $g=1 \times{\rm dim} \, \mathfrak{g}$. Furthermore, they also involve their first time derivative and then $e= 2 \times{\rm dim} \, \mathfrak{g}$. Using this information, $l= 3 \times {\rm dim}\, \mathfrak{g}$, and the map \eqref{map}, we have that the Chern-Simons theory possesses
\begin{eqnarray}
N_1 &=& e = 2\times {\rm dim}\, \mathfrak{g},\nonumber\\
N_2 &=& l + g - e = (3+1-2) \times {\rm dim}\, \mathfrak{g} \nonumber \\
&=& 2 \times {\rm dim}\, \mathfrak{g}
\end{eqnarray}
first- and second-class constraints in its Hamiltonian analysis, respectively. This is in agreement with Dirac's canonical analysis reported in Ref.~\onlinecite{Bla}. 

%\\\\\\\\\\\\\\\\\\\\\
\subsection{BF theory in four dimensions}
Now, we discuss BF theory in four dimensions based on $G$. As is well known, this is a reducible system in the context of Dirac's canonical formalism.\cite{Caicedo, Cuesta} Therefore, this system allows us to illustrate how the approach of this paper works in practice when there is reducibility. Furthermore, this theory is relevant because of its relation with Einstein's general relativity.\cite{Review} The theory is given by
\begin{eqnarray}
S[B, A]&=& \int_{M} B_i\wedge F^i, \label{BF}
\end{eqnarray}
where $F^i [A]= \d A^i + \frac12 f^i\,_{jk} A^j \wedge A^k \equiv \frac{1}{2} F^i_{\mu \nu} \d x^{\mu}\wedge \d x^{\nu}$ is the curvature of the ${\mathfrak g}$-valued connection 1-form $A^i$ and $B^i = \frac{1}{2} B^i_{\mu \nu} \d x^{\mu}\wedge \d x^{\nu}$ is a set of ${\mathfrak g}$-valued $2$-forms. The theory has $N=(4+6) \times{\rm dim} \, \mathfrak{g}= 10 \times{\rm dim} \, \mathfrak{g}$ field variables $A^i_{\mu}$ and $B^i_{\mu\nu}$. The variation of the action principle yields the following variational derivatives ${\mathcal E}^i$ and ${\mathcal E}'^i$
\begin{eqnarray} \label{eqbf}
\delta B^i &:&  {\mathcal E}^i:= F^i, %\Leftrightarrow \epsilon^{\mu\nu\gamma\delta} F^i_{\gamma\delta} = 0,
 \nonumber\\
\delta A^i &:& {\mathcal E}'^i:= -\mathcal{\mathcal{D}} B^i,
\end{eqnarray}
where $\mathcal{D} B^i =\d B^i+ f^i{}_{jk}A^j\wedge B^k$. Then, the equations of motion are
\begin{eqnarray}
{\mathcal E}^i =0, \qquad {\mathcal E}'^i = 0. \label{embf}
\end{eqnarray}
On the other hand, performing the spacetime decomposition of the fields, we get (modulo surface terms)
\begin{eqnarray}
S[B, A]&=& \frac{1}{2}\int_{M} \d^4x \, \epsilon^{0abc} \left( B_{i0a} F^{i}_{bc} +B_{iab}\dot{A}^i_c 
+ A^i_0 \mathcal{D}_c B_{iab} \right).\label{bfthe}
\end{eqnarray}
{\it Lagrangian constraints}. The points of the tangent bundle $T\cal{C}$ are locally labeled with
$20 \times{\rm dim} \, \mathfrak{g}$ variables given by $(A^i_\mu, B^i_{\mu\nu}, \dot{A}^i_\mu, \dot{B}^i_{\mu\nu})$. The corresponding symplectic structure (\ref{gp1}), energy (\ref{energy}), and the differential of the energy are
\begin{eqnarray}
\Omega&=& \frac{1}{2}\int \d^{3}x \, \epsilon^{0abc} \ed B^i_{ab} \ew \ed A_{ic} , \nonumber\\
E&=& -\frac{1}{2}\int \d^{3}x \, \epsilon^{0abc}\left(B_{i0a} F^{i}_{bc}+ A^i_0 \mathcal{D}_c B_{iab} \right) , \nonumber\\
\ed E&=& -\frac{1}{2}\int \d^{3}x \, \epsilon^{0abc}\left[ \mathcal{D}_c B_{iab}\, \ed A^i_0  -\left(  f_{ijk}A^i_0 B^k_{ab} +2 \mathcal{D}_c B^i_{0b} \right)\ed A^j_c \right. \nonumber \\
&&\left.  +F^i_{bc} \,\ed B_{i0a} -\mathcal{D}_c A^i_0 \, \ed B_{iab} \right].
\end{eqnarray}

A basis of $\ker \Omega$ is formed by the following $ (1+3+4+6) \times {\rm dim} \, \mathfrak{g} =14 \times {\rm dim} \, \mathfrak{g} $ null vectors $\{ Z_{1i}, Z^a_{2i}, Z_{3i}^{\mu}, Z^{\mu \nu}_{4i} $\}:
\begin{eqnarray}
 Z_{1i}&=&\int \d^{3}x \, \frac{\delta}{\delta A^i_0}, \qquad Z^a_{2i}=\int \d^{n-1}x \, \frac{\delta}{\delta B^i_{0a}}, \nonumber\\
 Z^{\mu}_{3 i}&=&\int \d^{3}x \, \frac{\delta }{\delta \dot{A}^i_{\mu}}, \qquad Z^{\mu \nu}_{4 i}=\int \d^{3}x \, \frac{\delta}{\delta \dot{B}^i_{\mu \nu}}, \label{onvbf}
\end{eqnarray}
but only $Z_{1i}$ and $Z^a_{2i}$ generate $(1+3) \times {\rm dim} \, \mathfrak{g} = 4  \times {\rm dim} \, \mathfrak{g}$ projectable Lagrangian constraints, given by
\begin{eqnarray}
\int \d^3x \, \varphi_{i}&:=& Z_{1i}\cdot \ed E = -\frac{1}{2}\int \d^3x \, \epsilon^{0abc}\mathcal{D}_a B_{ibc}\approx 0, \nonumber\\
\int \d^3x \, \varphi^a_i&:=& Z^a_{2i}\cdot \ed E = -\frac{1}{2}\int \d^3x \, \epsilon^{0abc} F_{ibc}\approx 0.
\end{eqnarray}
There are no more projectable constraints. However, the constraints $\varphi^a_i$ are not independent among themselves because they satisfy $\mathcal{D}_a \varphi^a_i=0$ due to Bianchi identities. These are $1 \times {\rm dim} \, \mathfrak{g}$ reducibility conditions of the constraints and so there are only $(3-1) \times{\rm dim} \, \mathfrak{g} = 2 \times{\rm dim} \, \mathfrak{g}$ independent constraints among $\varphi^a_i$.  Then, this theory possesses $3 \times {\rm dim} \, \mathfrak{g}$ independent projectable constraints.

Continuing with the theoretical framework of Sec. \ref{sec2}, one set of equations in Eq. (\ref{gp2+}) acquires the form
\begin{eqnarray}\label{news1}
\epsilon^{0abc} \alpha^i_c&=& \epsilon^{0abc}\mathcal{D}_c A^i_{0},\nonumber \\
\epsilon^{0abc} \alpha^i_{bc}&=& \epsilon^{0abc} \left( 2\mathcal{D}_b B^i_{0c}+f^i{}_{jk}A^k_0 B^j_{bc}\right).
\end{eqnarray}
Now, we substitute $\alpha^i_{a} = \dot{A}^i_{a}$ and $\alpha^i_{ab} = \dot{B}^i_{ab}$ into (\ref{news1}). This implies that the field variables must satisfy
the following $(3+3) \times {\rm dim} \, \mathfrak{g} = 6 \times {\rm dim} \, \mathfrak{g}$ non-projectable Lagrangian constraints
\begin{eqnarray}
\varphi^{ab}_i:= \epsilon^{abc0} F_{c0}^i \approx 0 ,\qquad \varphi^a_{2i}:= \epsilon^{\mu \nu \gamma a} \mathcal{D}_{\gamma} B^i_{\mu \nu} \approx 0.
\end{eqnarray}

In summary, we have $l=(10-1) \times {\rm dim} \, \mathfrak{g} = 9 \times {\rm dim} \, \mathfrak{g} $ independent constraints, because of the $1 \times{\rm dim} \, \mathfrak{g}$ reducibility conditions $\mathcal{D}_a \varphi^a_i=0$. Reducibility of the constraints must be taken into account in the counting of physical degrees of freedom, as was explained in the general theory, this is illustrated in what follows.

\subsubsection{Degree of freedom count}
{\it Associated vector fields}. It can be verified that the following $4  \times {\rm dim} \, \mathfrak{g}$ vector fields, modulo the null vectors \eqref{onvbf},
\begin{eqnarray}\label{BFav}
Z_5[N^i]&=& -\int \d^{3}x \left( \mathcal{D}_a N^i \frac{\delta}{\delta A^i_a} +f^i{}_{jk} B^j_{ab}N^k \frac{\delta}{\delta B^i_{ab}} \right), \nonumber \\
Z_6[M^i_a]&=&-2\int \d^{3}x \, \mathcal{D}_{[a} M^i_{b]} \frac{\delta}{\delta B^i_{ab}},
\end{eqnarray}
are the associated vectors to the projectable constraints $\varphi_i$  and $\varphi^a_i$, respectively. Actually, they satisfy $Z_5[N^i] \cdot\Omega= \ed\varphi_1 [N^i]$ and $Z_6[M^i_a] \cdot\Omega= \ed \varphi_2[M^i_a]$, where $\varphi_1[N^i]= \int \d^{3}x \, N^i \varphi_i$ and $\varphi_2[M^i_a]= \int \d^{3}x \, M^i_a \varphi^a_i$.

On the other hand, we have shown in the general theory of Sec. \ref{sec2} that reducibility is also reflected in the associated vectors. Let us illustrate this fact in this theory. Notice that $Z_6$, which is associated with the reducible constraints $\varphi^a_i$, has $3 \times{\rm dim} \, \mathfrak{g}$ parameters $M^i_a$, but taking into account the reducibility of the constraints ($\mathcal{D}_a \varphi^a_i=0$), we observe that, if we choose $M^i_a= \mathcal{D}_a P^i $ the vector field becomes
\begin{eqnarray}
Z_6 [\mathcal{D}_a P^i]&=&-2\int \d^{3}x \, \mathcal{D}_{[a} \mathcal{D}_{b]}P^i \frac{\delta}{\delta B^i_{ab}} =2\int \d^{3}x \, f^i{}_{jk} F^k_{ab}P^j \frac{\delta}{\delta B^i_{ab}},
\end{eqnarray} which vanishes under the inclusion, because of $\varphi^a_i$. This means that we have $1 \times{\rm dim} \, \mathfrak{g}$ reducibility conditions in the vector fields (we have $1 \times{\rm dim} \, \mathfrak{g}$ choices of the parameters $P^i$). Therefore, not all the parameters involved in $Z_6$ are independent, only $(3-1) \times{\rm dim} \, \mathfrak{g}= 2 \times{\rm dim} \, \mathfrak{g}$ of them are.

{\it Null associated vector fields}. Now, taking into account the original null vectors \eqref{onvbf} and the associated ones \eqref{BFav}, we form the vector fields
\begin{eqnarray}
Z_7&=& Z_{1i}[ Q^i]+Z^a_{2i}[Q^i_{2a}]+Z^{\mu}_{3i}[Q^i_{3\mu}]+ Z^{\mu \nu}_{4i}[Q^i_{\mu \nu}]+Z_5[N^i],\nonumber \\
Z_8&=& Z_{1i}[ R^i]+Z^a_{2i}[R^i_{2a}]+Z^{\mu}_{3i}[R^i_{3\mu}]+ Z^{\mu \nu}_{4i}[R^i_{\mu \nu}]+Z_6[M^i_a].
\end{eqnarray}
Demanding them to be tangent to the Lagrangian constraint surface fixes the following parameters
\begin{align}
&Q^{i} \, \, \, \approx -\mathcal{D}_0 N^i , && Q^{i}_{2a}\approx -f^i{}_{jk} B^j_{0a} N^k, \nonumber \\
&Q^{i}_{3a}\approx - \partial_0\left(  \mathcal{D}_a N^i\right), &&Q^i_{ab} \approx -\partial_0 \left(f^i{}_{jk} B^j_{ab}N^k\right),\nonumber\\
&R^{i}\, \, \, \approx 0 , && R^{i}_{2a} \approx -2\mathcal{D}_{[0} M^i_{a]}, \nonumber \\
&R^{i}_{3a} \approx 0, &&R^i_{ab} \approx -2\partial_0 \left( \mathcal{D}_{[a} M^i_{b]} \right),
\end{align}
where we have introduced the new parameters $M^i_0$ because of $B^i_{0a}=-B^i_{a0}$. Notice that the parameters $Q^{i}_{30}$, $Q^i_{0a}$, $R^{i}_{30}$, and $R^i_{0a}$ are left arbitrary because they smear the vector fields $ Z_{3i}^{0}$, and $Z_{4i}^{0a}$, which are tangent to the Lagrangian constraint surface. Therefore, we find that the null associated vector fields are
\begin{eqnarray}\label{nuasbf}
Z_7[N^i]&=&  -\int \d^{3}x \left[ \mathcal{D}_{\mu} N^i \frac{\delta}{\delta A^i_{\mu}} +f^i{}_{jk} B^j_{\mu \nu}N^k \frac{\delta}{\delta B^i_{\mu \nu}} + \partial_0\left( \mathcal{D}_{a} N^i \right) \frac{\delta}{\delta \dot{A}^i_{a}} \right. \nonumber\\
&&  \left.+\partial_0\left( f^i{}_{jk} B^j_{ab}N^k \right) \frac{\delta}{\delta \dot{B}^i_{ab}} \right] , \nonumber\\
Z_8[M^i_{\mu}]&=& -2 \int \d^{3}x \left[ \mathcal{D}_{[\mu} M^i_{\nu]} \frac{\delta}{\delta B^i_{\mu \nu}}  +\partial_0\left(  \mathcal{D}_{[a} M^i_{b]} \right) \frac{\delta}{\delta B^i_{ab}} \right].
\end{eqnarray}
Notice that the part that contains $M^i_0$ is not a new vector, it is of the form $-2 \int \d^{3}x \mathcal{D}_{[a} M^i_{0]} \frac{\delta}{\delta B^i_{0a}}$ which is generated by $Z^a_{2i}$. Therefore, there are only $(3-1) \times  {\rm dim} \, \mathfrak{g} = 2 \times {\rm dim} \, \mathfrak{g}$ independent vectors in $Z_8$.

{\it Degree of freedom count}. In sum, there are $B=(14+1 +2) \times{\rm dim} \, \mathfrak{g} = 17 \times{\rm dim} \, \mathfrak{g}$ independent null vector fields \eqref{onvbf} and \eqref{nuasbf}, and we have that there are $C=(1+2) \times  {\rm dim} \, \mathfrak{g} = 3 \times {\rm dim} \, \mathfrak{g}$ independent associated vectors in Eq. (\ref{BFav}) (or, equivalently, $3 \times {\rm dim} \, \mathfrak{g}$ linearly combinations of independent constraints used in the construction of the associated vector fields). Therefore, the number of physical degrees of freedom is locally 
\begin{eqnarray}
\frac{1}{2} \operatorname{Rank} \iota^* \Omega&=&10 \times {\rm dim}\, \mathfrak{g} -\frac{(17+3)}{2} \times {\rm dim}\, \mathfrak{g}=0.
\end{eqnarray}
as expected.\cite{Caicedo, Mondragon, Mond} Finally, using this result and Eq. (\ref{gl1}), we get that $\iota^* \Omega$ has $e+g= (2 \times 10 - 9) \times {\rm dim} \, \mathfrak{g} =11 \times {\rm dim} \, \mathfrak{g}$ null vectors. 

\subsubsection{Off-shell gauge transformations}
The gauge transformations are obtained as follows. First, we form the gauge vector 
\begin{eqnarray}
\tilde{X}&:=& Z^0_{3i}[P^i_{0}]+ Z^{0a}_{4i}[ P^i_{0a}]+Z_7[-\varepsilon^i]+ Z_8\left[-\varepsilon^i_{\mu} \right] \nonumber\\
&:=&\int \d^{3}x\left[ P^i_{0}\frac{\delta}{\delta \dot{A}^i_{0}} + P^i_{0a}\frac{\delta}{\delta \dot{B}^i_{0a}}+ \mathcal{D}_{\mu} \varepsilon^i \frac{\delta}{\delta A^i_{\mu}} +\partial_0 \left(\mathcal{D}_{a} \varepsilon^i \right) \frac{\delta}{\delta \dot{A}^i_{a}}\nonumber \right.\\
&& +\left(2\mathcal{D}_{\mu} \varepsilon^i_{\nu}+f^i{}_{jk} B^j_{\mu \nu}\varepsilon^k \right)\frac{\delta}{\delta B^i_{\mu \nu}}  \left.+\partial_0\left(2\mathcal{D}_{a} \varepsilon^i_{b}+f^i{}_{jk} B^j_{ab}\varepsilon^k \right)\frac{\delta}{\delta \dot{B}^i_{ab}} \right],
\end{eqnarray} where we have used $-\varepsilon^i$, and $-\varepsilon^i_{\mu}$ as smearing functions. Continuing with the procedure, the second step of Sec. \ref{GT} yields
\begin{equation}
P^i_{0} = \partial_0 \left(\mathcal{D}_{0}\varepsilon^i\right), \qquad
P^i_{0a}=\partial_0\left(2\mathcal{D}_{[0} \varepsilon^i_{a]}+f^i{}_{jk} B^j_{0a}\varepsilon^k \right).
\end{equation}
Hence, the gauge vector field acquires the form 
\begin{eqnarray}
\tilde{X}&=&\int \d^{3}x \left[ \mathcal{D}_{\mu} \varepsilon^i \frac{\delta}{\delta A^i_{\mu}} +\partial_0 \left(\mathcal{D}_{\mu} \varepsilon^i \right) \frac{\delta}{\delta \dot{A}^i_{\mu}} \right.+\left( 2\mathcal{D}_{[\mu} \varepsilon^i_{\nu]}+f^i{}_{jk} B^j_{\mu \nu}\varepsilon^k \right)\frac{\delta}{\delta B^i_{\mu \nu}} \nonumber \\
&& \left.+\partial_0\left(2\mathcal{D}_{[\mu} \varepsilon^i_{\nu]}+f^i{}_{jk} B^j_{\mu \nu}\varepsilon^k \right)\frac{\delta}{\delta \dot{B}^i_{\mu \nu}} \right].
\end{eqnarray}
Therefore, the gauge transformations are $\delta_{\varepsilon} A^i_{\mu}= \mathcal{D}_{\mu} \varepsilon^i$ and $
\delta_{\varepsilon} B^i_{\mu \nu}=2\mathcal{D}_{[\mu} \varepsilon^i_{\nu]}+f^i{}_{jk} B^j_{\mu \nu}\varepsilon^k$, which in terms of differential forms are
\begin{eqnarray}
\delta_{\varepsilon} A^i= \mathcal{D} \varepsilon^i, \qquad %\nonumber\\
\delta_{\varepsilon} B^i=\mathcal{D}\varepsilon'^i+f^i{}_{jk} B^j \varepsilon^k,\label{bftran}
\end{eqnarray}
where we have defined $\varepsilon'^i:= \varepsilon^i_{\mu} \d x^{\mu}$. The Noether identities  (from which these gauge transformations can also be read off) are given by $\mathcal{D} {\mathcal E}'^i- f^i\,_{jk}  B^ j\wedge {\mathcal E}^k =0$ and  $\mathcal{D} {\mathcal E}^i =0$. Furthermore, as is well-known the action \eqref{BF} is invariant under spacetime diffeomorphisms, only that diffeomorphism invariance is not independent from (\ref{bftran}). Indeed, using Cartan's formula \eqref{Cartan}, we have
\begin{eqnarray}
\pounds_{\xi} B^i&=& \mathcal{D} \left(\xi \cdot B^i \right)+ f^i{}_{jk}B^j \left( \xi \cdot A^k\right)+ \xi\cdot \mathcal{D}B^i, \nonumber\\
\pounds_{\xi} A^i &=& \mathcal{D} \left( \xi \cdot A^i \right)+ \xi \cdot F^i, \label{diffbf}
\end{eqnarray}
which are the transformations \eqref{bftran} with field-dependent parameters $\varepsilon^i=\xi \cdot A^i$ and $\varepsilon'^i=\xi \cdot B^i$ plus a trivial gauge transformation.

From \eqref{bftran} we can count the number of effective independent gauge parameters $e$ involved in the gauge transformations. Since it involves $\varepsilon^i$, $\dot{\varepsilon}^i$, $\varepsilon^i_0$, $\varepsilon^i_{a}$, and $\dot{\varepsilon}^i_{a}$, but we have the reducibility condition on the gauge parameters $\varepsilon^i_{a}$, then the count is as follows: $e= (1\times{\rm dim} \, \mathfrak{g})+ (1\times{\rm dim} \, \mathfrak{g}) + (1\times{\rm dim} \, \mathfrak{g})+ (2\times{\rm dim} \, \mathfrak{g})+ (2\times{\rm dim} \, \mathfrak{g})$, respectively. Hence, we have $g=4 \times{\rm dim} \, \mathfrak{g}$. Therefore, according to the map (\ref{map}), the BF theory must have
\begin{eqnarray}
N_1&=& e= 7 \times {\rm dim} \,\mathfrak{g},\nonumber \\
N_2&=& l+g-e= (9+4 -7) \times {\rm dim} \,  \mathfrak{g}\nonumber\\
&=&  6 \times {\rm dim} \,  \mathfrak{g}
\end{eqnarray}
first- and second-class constraints, respectively, if its Dirac's canonical analysis is performed.

%\\\\\\\\\\\\\\\\\\\\\\\\\\\\\\
\subsection{General relativity}

Finally, in this subsection, we present the Lagrangian analysis for first-order general relativity. Our starting point is the Palatini action with cosmological constant $\Lambda$ given by
\begin{eqnarray}
S[e, \omega]= \alpha  \int_{M} \left[  e^I\wedge e^J\wedge *R_{IJ}[\omega] -\frac{\Lambda}{6}  e^I\wedge e^J  \wedge *\left(e_{I}\wedge e_{J}\right) \right], \label{GR}
\end{eqnarray}
where $e^I$ is an orthonormal frame of 1-forms; the Lorentz indices $I, J = 0, 1, 2, 3$ are raised and lowered with the Minkowski metric $(\eta_{IJ}) = \mbox{diag}(-1,+1,+1,+1)$, $\omega^I\,_J$ is a Lorentz connection 1-form, $\omega_{IJ}=-\omega_{JI}$, and $R^I\,_J= \d \omega^I\,_J + \omega^I\,_K \wedge \omega^K\,_J$ ($=\frac 12 R^I\,_{JKL} e^K \wedge e^L$) is its curvature. The definition of the (internal) dual operator is $ *v^{IJ}= \frac{1}{2}\epsilon^{IJKL} v_{KL}$ with $\epsilon_{0123}=1$. The constant $\alpha$ in front of the action is introduced for units. Notice that there are $N= 16+ 24$ field variables in $e^I_{\mu}$ and $\omega^{IJ}\,_{\mu}$, respectively. 

The variation of the action \eqref{GR} is 
\begin{eqnarray}
\delta S [e,\omega] &=& \int_M \left[ {\mathcal E}_I \wedge \delta e^I  + {\mathcal E}_{IJ} \wedge \delta \omega^{IJ} \right]  + \int_{\partial M} \alpha \ast \left ( e_I \wedge e_J \right ) \wedge \delta \omega^{IJ} ,
\end{eqnarray}
where ${\mathcal E}_I$ and ${\mathcal E}_{IJ}$ are the variational derivatives, given by
\begin{eqnarray}\label{cortas}
{\mathcal E}_I &:=& - 2 \alpha e^J \wedge \ast \left ( R_{IJ} - \frac{\Lambda}{3} e_I \wedge e_J \right ), \nonumber \\
{\mathcal E}_{IJ} &:=& - \alpha D \ast \left ( e_I \wedge e_J \right ), 
\end{eqnarray}
and $D \ast \left ( e_I \wedge e_J \right ) = \d \ast \left ( e_I \wedge e_J \right ) - \omega^K\,_I \wedge \ast \left ( e_K \wedge e_J \right ) - \omega^K\,_J  \wedge \ast \left ( e_I \wedge e_K \right )  $
and so the equations of motion are
\begin{eqnarray}
{\mathcal E}_I =0, \qquad {\mathcal E}_{IJ}=0. \label{emrg}
\end{eqnarray}

Now, we start our Lagrangian analysis by performing the spacetime decomposition of the fields. The Palatini action acquires the form (modulo surface terms)
\begin{eqnarray}
S[e, \omega] &= & 
\alpha \int \d^4x \, \epsilon^{0abc} \left[ e^I_0 e^J_a *R_{IJbc}  + * \left( e_{Ia} e_{Jb}\right) \dot{\omega}^{IJ}{}_c  +\omega^{IJ}{}_0 \mathcal{D}_c *\left(e_{Ia} e_{Jb}\right)\right. \nonumber\\
&&  \left.  -\frac{2\Lambda}{3} e^I_0 e^J_a *\left( e_{Ib} e_{Jc}\right) \right] . 
\end{eqnarray}
{\it Lagrangian constraints}.  In this case, the points of the tangent bundle $T\cal{C}$ are locally labeled with
$80$ variables given by $(e^I_{\mu}, \omega^{IJ}\,_{\mu}, \dot{e}^I_{\mu}, \dot{\omega}^{IJ}\,_{\mu})$. The corresponding symplectic structure (\ref{gp1}), energy (\ref{energy}), and the differential of the energy are
\begin{eqnarray}\label{ssgr}
\Omega&=& \alpha \int \d^{3}x \, \epsilon_{IJKL} \epsilon^{0abc} e^{I}_{a} \ed e^{J}_{b} \ew \ed \omega^{KL}{}_c, \nonumber\\
E&=&-\alpha \int \d^{3}x \, \epsilon^{0abc} \left[ e^I_{0} e^J_{a} *R_{IJbc} + \omega_{IJ0} D_{c} *\left( e^I_{a} e^J_{b}\right) -\frac{2\Lambda}{3}  e^I_0 e^J_a *\left( e_{Ib} e_{Jc}\right) \right], \nonumber\\
\ed E&=& - \alpha  \int \d^{3}x \,\epsilon^{0abc} \left\{ -\left[ e^J_0 *R_{IJbc} + \epsilon_{IJKL}e^J_b D_c \omega^{KL}{}_0  +2\Lambda e^J_0 *\left( e_{Ib} e_{Jc} \right) \right] \ed e^I_a \right.\nonumber\\
&& +\left[ 2 D_c *\left( e_{I0} e_{Jb}\right) +\omega_{IK0} *\left( e_{Jb} e^K_c \right)  -\omega_{JK0} *\left( e_{Ib} e^K_c \right)  \right] \ed \omega^{IJ}{}_{a} \nonumber\\
&&\left. +\left[ e^J_{a} *R_{IJbc} -\frac{2\Lambda}{3} e^J_a *\left( e_{Ib} e_{Jc} \right) \right] \, \ed e^I_{0} +D_{c} *\left( e_{Ia} e_{Jb}\right) \ed \omega^{IJ}{}_{0} \right\}.
\end{eqnarray}
A basis of $\ker \Omega$ is given by the $4+6+6+16+24=56$ null vector fields $\{ Z_{I}, Z_{IJ}, Z^{(bc)} , Z^{\mu}_{I}, Z^{\mu}_{IJ} \}$,
\begin{align}\label{noGR}
Z_{I}&=\int \d^{3}x \, \frac{\delta}{\delta e^{I}_{0} }, &Z_{IJ}&=\int \d^{3}x \, \frac{\delta}{\delta {\omega}^{IJ}{}_{0}}, \nonumber\\ %Z^i_{1}&=\int \d^{3}x \, \eta^{iIJ}_{1a} \frac{\delta}{ \delta \omega^{IJ}_a}, &Z^i_{2}&=\int \d^{3}x \, \eta^{iIJ}_{2a} \frac{\delta}{ \delta \omega^{IJ}_a}
Z^{(bc)}&= \int \d^{3}x \, \delta^{(b}_a\epsilon^{c)0de} e^I_{d} e^J_{e} \frac{\delta}{ \delta \omega^{IJ}{}_a}, &
 Z^{\mu}_{I}&=\int \d^{3}x \, \frac{\delta}{\delta \dot{e}^{I}_{\mu} }, \nonumber\\
 Z^{\mu}_{IJ}&=\int \d^{3}x \, \frac{\delta}{\delta \dot{\omega}^{IJ}{}_{\mu}}.
\end{align}
However, from this set, only $ \{ Z_{I}, Z_{IJ}, Z^{(bc)}\}$ generate, respectively, the following $4+6+6=16$ projectable Lagrangian constraints:
\begin{eqnarray}
\varphi_I&:=& -\alpha\epsilon^{0abc}\left[ e^J_a *R_{IJbc}-\frac{2\Lambda}{3} e^J_a *\left( e_{Ib} e_{Jc} \right) \right] \approx 0, \nonumber \\
\varphi_{IJ}&:=&- \alpha\epsilon^{0abc} D_{c} *\left( e_{Ia} e_{Jb} \right) \approx 0, \nonumber\\
\varphi^{(bc)}&:=& \delta^{(b}_a\epsilon^{c)0de} e^I_{d} e^J_{e}  C^a_{IJ} \approx 0,\label{onvgr}
\end{eqnarray}
where 
\begin{eqnarray}
C^a_{IJ}&=&\alpha \epsilon^{0abc} \left[2 D_c *\left( e_{I0} e_{Jb}\right) +\omega_{IS0} *\left( e_{Jb} e^S_c \right)  -\omega_{JS0} *\left( e_{Ib} e^S_c \right)  \right].
\end{eqnarray}
There are no additional constraints. Notice that the constraints $\varphi_I$ and $\varphi_{IJ}$ are part of the equations of motion, and that the  constraints $\varphi^{(bc)}$ are not. 

Now, we search for the non-projectable constraints. Equation \eqref{gp2+} becomes
\begin{eqnarray}
- \alpha \epsilon^{0abc} \epsilon_{IJKL} e^J_b \alpha^{KL}{}_c&=& -\alpha \epsilon^{0abc}\left[ e^J_0 *R_{IJbc} + \epsilon_{IJKL}e^J_b D_c \omega^{KL}{}_0  +2\Lambda e^J_0 *\left( e_{Ib} e_{Jc} \right) \right], \nonumber\\
\alpha \epsilon^{0abc} \epsilon_{IJKL} e^K_b \alpha^L_c&=& C^a_{IJ}, \nonumber\\
0&=& \varphi_I ,\nonumber\\
0&=& \varphi_{IJ}.
\end{eqnarray}
By substituting $\alpha^I_a=\dot{e}^{I}_{a}$ and $\alpha^{IJ}{}_a=\dot{\omega}^{IJ}{}_{a}$, in these equations we get the following $12+18=30$ non-projectable constraints: 
\begin{eqnarray}
\varphi^a_I &:=& \epsilon^{a \beta \gamma \delta}\left[ \alpha e^J_{\beta} *R_{IJ\gamma\delta}-\frac{2\Lambda}{3} e^J_{\beta} *\left(e_{I\gamma}e_{J\delta} \right) \right] \approx 0, \nonumber\\
\varphi^a_{IJ}&:=& \alpha \epsilon^{a \beta \gamma \delta} D_{\delta} *\left( e_{I\beta} e_{J\gamma}\right) \approx 0.
\end{eqnarray}
Therefore, the total number of Lagrangian constraints is  $l=16+30= 46$.

\subsubsection{Degree of freedom count}
{\it Associated vector fields}. Here, we only write the smeared projectable constraints that generate associated vector fields,
\begin{eqnarray}\label{pcsgr}
\varphi[N^I]= \int \d^{3}x \,  N^I \varphi_I, \qquad \varphi[M^{IJ}]= \int \d^{3}x \, M^{IJ} \varphi_{IJ}.
\end{eqnarray}
Their corresponding associated vector fields, modulo the null vector fields \eqref{noGR}, are
\begin{eqnarray}
Z_1[N^I]&=& -\int \d^{3}x \left[  \left( D_a N^I \right. + \Theta^{JK}_b \,{}^{I}_{a} \epsilon^{0bcd} \epsilon_{JKLM}N^{L} D_d e^M_c \right)\frac{\delta}{\delta e^I_a}  \nonumber\\
&& + \left( \frac{ \epsilon^{IJKL}}{2}*R_{MKLN} N^M e^{N}_{a}  \left. -  \frac{ *R*_{MN}{}^{IJ}}{2} N^M e^{N}_{a}   +\Lambda N^{[I}e^{J]}_a\right) \frac{\delta}{\delta \omega^{IJ}{}_a} \right],\nonumber\\
Z_2[M^{IJ}]&=& \int \d^{3}x \left( M^I{}_{J} e^J_{a} \frac{\delta}{\delta e^I_a} - D_a M^{IJ}\frac{\delta}{\delta \omega^{IJ}{}_a}\right),
\label{nullvGR}
\end{eqnarray}
with $\Theta^{JK}_b \,{}^{I}_{a}$ being the right inverse of $\Omega^{a}_{I} \,  {}^{b}_{JK}= \epsilon_{IJKL}\epsilon^{0abc} e^L_c$, i.e., it satisfies $\Omega^{a}_{I} \,  {}^{b}_{JK}  \Theta^{JK}_b \,{}^{L}_{c} = \delta^a_c \delta^L_I$ (see Refs.~\onlinecite{Contreras, Contreras2} for details). 
Notice that the term that involves $\Theta^{JK}_b \,{}^{I}_{a}$ is proportional to $D_d e^P_c $, which vanishes because of the constraints \eqref{onvgr}. This term vanishes when we construct the null associated vector fields and for this reason the explicit form of $\Theta^{JK}_b \,{}^{I}_{a}$  is irrelevant in the computation. 

{\it Null associated vector fields}. Now, we need to check if the vectors \eqref{nullvGR} together with the original null vectors \eqref{noGR} of the symplectic structure are tangent to the Lagrangian constraint surface. Therefore, we form the following vectors:
\begin{eqnarray}
Z_3 &=& Z_1[N^I]+Z_{I}[P^I]+ Z_{IJ}[P^{IJ}]+Z^{(bc)}[P_{bc}] + Z^{\mu}_{I}[P^I_{\mu}] + Z^{\mu}_{IJ}[P^{IJ}_{\mu}] ,\nonumber\\
Z_4 &=& Z_2[M^{IJ}]+ Z_{I}[Q^I]+ Z_{IJ}[Q^{IJ}]+Z^{(bc)}[Q_{bc}] + Z^{\mu}_{I}[Q^I_{\mu}]  + Z^{\mu}_{IJ}[Q^{IJ}_{\mu}].
\end{eqnarray}
Next, we demand them to be tangent to the Lagrangian constraint surface, i.e., they have to override all the Lagrangian constraints. This requirement yields
\begin{align}
P^{IJ}&\approx \frac{1}{2}\left(-\epsilon^{IJKL}*R_{MKLN}  + *R*_{MN}{}^{IJ}  \right) N^M e^{N}_{0}  -\Lambda N^{[I} e^{J]}_0,\nonumber\\
P^{IJ}_{a} &\approx -\partial_0 \left[ \frac{1}{2}\left( \epsilon^{IJKL}*R_{MKLN} -   *R*_{MN}{}^{IJ} \right)N^M e^{N}_{a}  +\Lambda N^{[I}e^{J]}_a \right] ,  \nonumber\\
0&\approx\Theta^{JK}_b \,{}^{I}_{a} \epsilon^{0bcd} \epsilon_{JKLM}N^{L} D_d e^M_c,\nonumber\\
P^I& \approx  - D_0 N^I, \qquad \quad \quad\, P^I_{a} \approx -\partial_0 \left( D_{a} N^I\right),\nonumber\\
P_{bc}&\approx0, \qquad \qquad \qquad \quad\! Q_{bc} \approx 0,\nonumber\\
Q^I& \approx M^I{}_J e^J_0 , \qquad \quad\quad \, Q^{IJ}\approx - D_0 M^{IJ},  \nonumber\\
Q^I_{a}& \approx \partial_0 \left(  M^I{}_J e^J_{a} \right),\quad \quad Q^{IJ}_{a} \approx - \partial_0 \left( D_{a} M^{IJ} \right) .
\end{align}
Notice that in the equation $0 \approx \Theta^{JK}_b \,{}^{I}_{a} \epsilon^{0bcd} \epsilon_{JKLM}N^{L} D_d e^M_c$ there is nothing new. Actually, it is just consequence of the constraints \eqref{onvgr}. Therefore, the null associated vector fields are
\begin{eqnarray}\label{nulosdeverdad}
Z_3[N^I]&=& -\int \d^{3}x  \left[  D_{\mu} N^I  \frac{\delta}{\delta e^I_{\mu}} +\partial_0\left(D_{a} N^I \right)\frac{\delta}{\delta \dot{e}^I_{a}} \right. \nonumber\\
&&+\left( \frac{ \epsilon^{IJKL}}{2}*R_{MKLN}  N^M e^{N}_{\mu} -  \frac{ *R*_{MN}{}^{IJ}}{2} N^M e^{N}_{\mu}   +\Lambda N^{[I}e^{J]}_{\mu}\right) \frac{\delta}{\delta \omega^{IJ}{}_{\mu}} \nonumber\\
&&+ \partial_0 \left( \frac{ \epsilon^{IJKL}}{2}*R_{MKLN} N^M e^{N}_{a} \left. -  \frac{ *R*_{MN}{}^{IJ}}{2} N^M e^{N}_{a} +\Lambda N^{[I}e^{J]}_{a}\right) \frac{\delta}{\delta \dot{\omega}^{IJ}{}_{a}} \right],\nonumber\\
Z_4[M^{IJ}]&=&  \int \d^{3}x \left[ M^I{}_J e^J_{\mu} \frac{\delta}{\delta e^I_{\mu}} - D_{\mu}M^{IJ} \frac{\delta}{\delta \omega^{IJ}{}_{\mu}} + \partial_0 \left( M^I{}_J e^J_{a} \right) \frac{\delta}{\delta \dot{e}^I_{a}} - \partial_0 \left(D_{a}M^{IJ} \right) \frac{\delta}{\delta \dot{\omega}^{IJ}{}_{a}} \right].\nonumber\\
\label{gtgr}
\end{eqnarray}
Notice that the parameters  $P^{IJ}_0, P^I_0, Q^I_0,$ and $Q^{IJ}_0$ are left arbitrary. This is so because the vector fields  $Z^{0}_{IJ}$ and $Z^0_I$ are tangent to the Lagrangian constraint surface.

{\it Degree of freedom count}. We are ready to make the counting of the physical degrees of freedom. There are $N=40$ field variables in $e^I_{\mu}$ and $\omega^{IJ}\,_{\mu}$, there are $B=56+10=66$ null vectors given by (\ref{noGR}) and (\ref{nulosdeverdad}), and  the number of linearly independent combinations of constraints \eqref{pcsgr} [to get the associated vector fields \eqref{nullvGR}] is $C=10$. Therefore, the number of the physical degrees of freedom is locally
\begin{equation}
\frac{1}{2} {\rm Rank} \, \iota^*\Omega = 40-\frac{1}{2} \left(66+10 \right)=2,
\end{equation}
as expected. 

From this result (and the fact that $N=40$ and $l=46$), Eq. (\ref{gp1}) implies $g+e=30$ null vectors for $\iota^* \Omega$.

\subsubsection{Off-shell gauge transformations}
Now, we look for the gauge transformations, which we have split into two parts in order to enhance the exposition. We begin with local Lorentz transformations. Following the steps explained in Sec. \ref{GT}, first, we use the null associated vector $Z_4[\varepsilon^{IJ}]$, $Z^{0}_{IJ}$ and $Z^0_I$ and we form the vector field ($\varepsilon^{IJ}=-\varepsilon^{JI}$)
\begin{eqnarray}
\tilde{X}_1&:=&  Z_4[\varepsilon^{IJ}]+ Z^{0}_{IJ}[Q^{IJ}_{0}]+ Z^{0}_{I}[Q^{I}_{0}].\label{gauRG}
\end{eqnarray}
Second, we demand $\tilde{X}_1$ to be of the form
\begin{eqnarray} \label{gvfgr}
\tilde{X}&=&\int \d^{3}x\left[ \delta_{\varepsilon} e^I_{\mu}\frac{\delta}{\delta e^I_{\mu}}+\delta_{\varepsilon} \omega^{IJ}{}_{\mu}\frac{\delta}{\delta \omega^{IJ}{}_{\mu}}+\partial_0(\delta_{\varepsilon} e^I_{\mu}) \frac{\delta}{\delta \dot{e}^I_{\mu}}  + \partial_0 \left( \delta_{\varepsilon} \omega^{IJ}{}_{\mu}\right)\frac{\delta}{\delta \dot{\omega}^{IJ}{}_{\mu}} \right],
\end{eqnarray}
which gives the following relations between the parameters $Q^{IJ}_{0}=\partial_0\left(\mathcal{D}_{0} \varepsilon^{IJ}  \right)$ and $Q^I_0= \partial_0 \left( M^I{}_J e^J_0 \right)$. Consequently, the gauge vector field reduces to 
\begin{eqnarray}
\tilde{X}_1&=&\int \d^{3}x  \left[ \varepsilon^I{}_J e^J_{\mu} \frac{\delta}{\delta e^I_{\mu}} - D_{\mu}\varepsilon^{IJ} \frac{\delta}{\delta \omega^{IJ}{}_{\mu}}  + \partial_0 \left( \varepsilon^I{}_J e^J_{\mu} \right) \frac{\delta}{\delta \dot{e}^I_{\mu}}- \partial_0 \left(D_{\mu}\varepsilon^{IJ} \right) \frac{\delta}{\delta \dot{\omega}^{IJ}{}_{\mu}} \right].
\end{eqnarray}
From this, we can read off the local Lorentz transformations
\begin{eqnarray}
\delta_{\varepsilon} e^I = \varepsilon^I\,_J e^J, \qquad
\delta_{\varepsilon} \omega^{IJ} = - D \varepsilon^{IJ}.
\end{eqnarray}

On the other hand, using the vectors $Z_3[-\varepsilon^I]$, $Z^{0}_{IJ}$, and $Z^{0}_{I}$, we form the vector $\tilde{X}_2:= Z_3[-\varepsilon^{I}]+Z^{0}_{IJ}[P^{IJ}_{0}]+Z^{0}_{I}[P^{I}_{0}]$, and demanding it to be of the form \eqref{gvfgr} gives
\begin{eqnarray}
P^{IJ}_0&=& \partial_0 \left[\frac{1}{2} \left( \epsilon^{IJKL}*R_{MKLN}  -  *R*_{MN}{}^{IJ}\right) \varepsilon^M e^{N}_{0}  +\Lambda \varepsilon^{[I}e^{J]}_0\right],\nonumber\\
P^I_0&=&  \partial_0 \left( D_0 \varepsilon^I\right),
\end{eqnarray}
and therefore
\begin{eqnarray}
\tilde{X}_2&=&\int \d^{3}x \left\{ D_{\mu} \varepsilon^I \frac{\delta}{\delta e^I_{\mu}} +\partial_0\left(D_{\mu} \varepsilon^I \right)\frac{\delta}{\delta \dot{e}^I_{\mu}}\right. \nonumber\\
&&+\left[ \frac{1}{2}\left( \epsilon^{IJKL}*R_{MKLN} - *R*_{MN}{}^{IJ}\right) \varepsilon^M e^{N}_{\mu} +\Lambda \varepsilon^{[I}e^{J]}_{\mu} \right] \frac{\delta}{\delta \omega^{IJ}{}_{\mu}} \nonumber\\
&&  \left.+ \partial_0 \left[ \frac{1}{2}\left( \epsilon^{IJKL}*R_{MKLN}-  *R*_{MN}{}^{IJ} \right) \varepsilon^M e^{N}_{\mu} +\Lambda \varepsilon^{[I}e^{J]}_{\mu} \right] \frac{\delta}{\delta \dot{\omega}^{IJ}{}_{\mu}} \right\}.
\end{eqnarray}
From this we can read off the gauge transformations \cite{Celada}
\begin{eqnarray}\label{tres}
\delta_{\varepsilon} e^I &=& D \varepsilon^I, \nonumber\\
\delta_{\varepsilon} \omega^{IJ} &=& \frac12 ( \epsilon^{IJKL} \ast R_{MKLN} - \ast R \ast{}_{MN}{}^{IJ}) \varepsilon^M e^N  + \frac{\Lambda}{2} \left ( \varepsilon^I e^J - \varepsilon^J e^I \right ).
\end{eqnarray}
A detailed analysis of Noether's identities (from which these gauge transformations come from) is reported in Ref.~\onlinecite{Celada}. Furthermore, the relationship between the local gauge transformations (\ref{tres}) and the diffeomorphism transformation of $e^I$ and $\omega^{IJ}$, using the Lie derivative as in previous examples, has also been explained in Ref.~\onlinecite{Celada}. 

There is an alternative way of showing this relationship based on the current approach, which we explain now. First, let us replace the Lorentz index in $\varphi_I$ by a spacetime index by defining $\varphi_{\mu}:= e^I_{\mu} \varphi_I$. With $\varphi_{\mu}$ and the constraints $\varphi_{IJ}$, we define the new constraints
\begin{eqnarray}
\mathcal{H}_{\mu}&:=&\varphi_{\mu}-\omega^{IJ}{}_{\mu} \varphi_{IJ}.
\end{eqnarray}
In order to identify the vector field associated with these constraints it is convenient to analyze them separately. We start by analyzing the constraint $\mathcal{H}_a$, and then $\mathcal{H}:= \mathcal{H}_0$. Notice that the constrains $\varphi_a$ can be written as $\varphi_a= -2 \epsilon^{0bcd}\epsilon_{IJKL} e^I_b e^J_c R^{KL}{}_{ad}$ where the identity $ \epsilon^{\alpha \beta \gamma \delta}\epsilon_{IJKL} e^K_{\gamma} e^L_{\delta}= 4e e^{[\alpha}_I e^{\beta]}_J$ was used twice. Therefore, we have that
\begin{eqnarray}
\mathcal{H}_a&=&\varphi_a-\omega^{IJ}{}_a \varphi_{IJ} \nonumber\\
&=&\partial_d\left[ \alpha\epsilon^{0bcd} *\left(e_{Ib} e_{Jc}\right) \omega^{IJ}{}_a \right]- \alpha\epsilon^{0bcd} *\left(e_{Ib} e_{Jc}\right)\partial_a \omega^{IJ}{}_d.
\end{eqnarray}
Defining the smeared constraints $\mathcal{H}[\xi^a]= \int \d^{3}x \, \xi^a \mathcal{H}_a$, it can be proved that it generates the associated vector field
\begin{eqnarray}
Z_5[\xi^a]&=& \int \d^{3}x \left[\left(\xi^b \partial_b e^I_a +e^I_b \partial_a \xi^b \right)\frac{\delta}{\delta e^I_a} + \left( \xi^b \partial_b \omega^{IJ}{}_a +\omega^{IJ}{}_b \partial_a \xi^b \right) \frac{\delta}{\delta \omega^{IJ}{}_a}\right] . 
\end{eqnarray}
Notice that the contribution of this vector to the gauge transformations is the so-called spatial diffeomorphism. In this sense, we can identify $\varphi_a$ as the Lagrangian version of the vector constraint.

On the other hand, from Dirac's canonical analysis, it is known that the symmetry corresponding to the projection on the cotangent bundle of the constraint $\mathcal{H}$, can be associated on-shell with diffeomorphisms that are normal to the spatial surface of the foliation. Let us show that this is also true in the current Lagrangian theoretical framework. We define the smeared constraint
\begin{equation}
\mathcal{H}[\xi^0]= \int \d^{3}x \, \xi^0 \mathcal{H}.
\end{equation}
We calculate its differential, and we obtain
\begin{eqnarray}
\ed \mathcal{H}[\xi^0]&=&\int \d^{3}x \left\{ \xi^0\left( \varphi_I \, \ed e^I_0-\varphi_{IJ}\, \ed \omega^{IJ}{}_0+\varphi^a_I\, \ed e^I_a +\varphi^a_{IJ} \, \ed \omega^{IJ}{}_a \right) \right. \nonumber\\ 
&& + \alpha\epsilon^{0abc}\epsilon_{IJKL} e^I_a\left[ \left( \xi^0 \partial_0 e^J_b +e^J_0\partial_b \xi^0\right)\,\ed \omega^{KL}{}_c \right. \nonumber\\
&&\left.  -\left( \xi^0 \partial_0 \omega^{KL}{}_c   \left.+\omega^{KL}{}_0\partial_c \xi^0\right) \ed e^J_b \right] \right\},
\end{eqnarray}
which does not possess an associated vector field, because it has components along $\ed e^I_0$ and $\ed \omega^{IJ}_0$. However, if we restrict ourselves to the Lagrangian constraint surface, the differential is reduced to
\begin{eqnarray}
\iota^* \left( \ed \mathcal{H}[\xi^0] \right)&=&\int \d^{3}x \, \alpha\epsilon^{0abc}\epsilon_{IJKL} e^I_a\left[ \left( \xi^0 \partial_0 e^J_b +e^J_0\partial_b \xi^0\right)\,\ed \omega^{KL}{}_c \right. \nonumber\\
&& -\left( \xi^0 \partial_0 \omega^{KL}{}_c  \left.+\omega^{KL}{}_0\partial_c \xi^0\right) \ed e^J_b \right] ,
\end{eqnarray}
and in that case, its associated vector field is
\begin{eqnarray}
Z_6[\xi^0]&=& \int \d^{3}x \left[\left(\xi^0 \partial_0 e^I_a +e^I_0 \partial_a \xi^0 \right)\frac{\delta}{\delta e^I_a} + \left( \xi^0 \partial_0 \omega^{IJ}{}_a +\omega^{IJ}{}_0 \partial_a \xi^0 \right) \frac{\delta}{\delta \omega^{IJ}{}_a} \right]. 
\end{eqnarray}
%The reason that we need to restrict ourselves to the constraints surface is because as we have shown, the diffeomorphism contains trivial parts.
Therefore, $Z_5$ and $Z_6$ can be combined and give rise to the vector field
\begin{equation}
Z_7[\xi^{\mu}]= \int \d^{3}x \left(\pounds_{\xi} e^I_a\frac{\delta}{\delta e^I_a}+ \pounds_{\xi} \omega^{IJ}{}_a \frac{\delta}{\delta \omega^{IJ}{}_a} \right), \label{vdiff}
\end{equation}
where $\pounds_{\xi} e^I_a = \xi^{\mu} \partial_{\mu} e^I_a +e^I_{\mu} \partial_a \xi^{\mu}$ and $\pounds_{\xi} \omega^{IJ}{}_a = \xi^{\mu} \partial_{\mu} \omega^{IJ}{}_a +\omega^{IJ}{}_{\mu} \partial_a \xi^{\mu}$. Then, we have the diffeomorphism transformations of $e^I_a$ and $\omega^{IJ}_a$. To obtain the diffeomorphism transformations of $e^I_0$ and $\omega^{IJ}{}_0$, we need to follow the steps described in the general theory of Sec. \ref{sec2}. That is, first we must form a new vector using $Z_7$ and the original null vectors \eqref{noGR}, and we must demand this new vector to be tangent to the Lagrangian constraint surface, which gives us the null associated vector field. Second, following the two steps described in Subsection \ref{GT} to form the vector field $\tilde{X}$ we obtain
\begin{eqnarray}
\tilde{X}&=& \int \d^{3}x \left[ \pounds_{\xi} e^I_{\nu} \frac{\delta}{\delta e^I_{\nu}} +\pounds_{\xi} \omega^{IJ}{}_{\nu} \frac{\delta}{\delta \omega^{IJ}{}_{\nu}}  +\partial_0 \left(\pounds_{\xi} e^I_{\nu} \right) \frac{\delta}{\delta \dot{e}^I_{\nu}}  + \partial_0 \left(\pounds_{\xi} \omega^{IJ}{}_{\nu} \right) \frac{\delta}{\delta \dot{\omega}^{IJ}{}_{\nu}} \right],
\end{eqnarray}
from which we can read the diffeomorphism transformations
\begin{eqnarray}\label{dos}
\delta_{\xi} e^I = \pounds_{\xi} e^I, \quad \delta_{\xi} \omega^{IJ} = \pounds_{\xi} \omega^{IJ}.
\end{eqnarray}

Finally, in order to use the map \eqref{map}, we have that the full gauge transformations involve the gauge parameters $\varepsilon^{IJ},\, \varepsilon^I$ and their  first (time) derivatives, so $g=10$ and $e=20$. Therefore, the  number of first- and second-class constraints that should appear in Dirac's canonical analysis is
\begin{eqnarray}
N_1 &=& e = 20 ,\nonumber\\
N_2 &=& l + g - e = 46+10-20=36.
\end{eqnarray}

%\\\\\\\\\\\\\\\\\\\\\\\Section IV\\\\\\\\\\\\\\\\\\\\\\\\
\section{Concluding remarks}\label{sec4}

We have generalized to field theory the geometric Lagrangian approach to the physical degree of freedom count reported in Ref.~\onlinecite{diaz} for point particle systems. We conclude this paper by making some comments: 

(a) We emphasize that in order to count the physical degrees of freedom of any Lagrangian theory under study using the current geometric Lagrangian approach we require only  knowing the Lagrangian constraints and the number of null vector fields of the symplectic structure (the original null vectors plus the null associated ones). This is essentially the result expressed in Eq. (\ref{pdfc}). Therefore, this approach does {\it not} require knowing the gauge transformations for the counting.

(b) Regarding the gauge transformations, we have taken one step forward from Ref.~\onlinecite{diaz}. We have shown that off-shell gauge transformations can be obtained using the current geometric Lagrangian approach, generalizing the previous results, where the gauge transformations were considered on-shell only. Furthermore, from the knowledge of the off-shell gauge transformations, we can determine the Lagrangian parameters $e$ and $g$.

(c) Because of item (b) and thanks to the map (\ref{map}), we can know the number of first- and second-class constraints of the corresponding Dirac's canonical analysis without having to do it. This is an important fact because the classification of the Hamiltonian constraints is usually a non-trivial task in the canonical analysis.

(d) We have given a detailed explanation of how we must deal with reducible systems, showing that the reducibility of the constraints is also reflected in their associated vector fields. Reducibility has to be taken into account by considering independent quantities in the counting only. This fact is also important when using the map between the Lagrangian and the Hamiltonian parameters. %Therefore, we have an alternative way to analyze singular field theories from the Lagrangian point of view which includes all kinds of singular systems. 

(e) In order to illustrate the approach, we have analyzed some relevant physical theories, which include BF theory and general relativity, showing that our geometric Lagrangian approach gives the correct information of these theories and that it is in agreement with the information we get from the corresponding Hamiltonian analyses. 

(f) Although we have focused on the geometric Lagrangian approach, it is worth mentioning that once we have shown how reducibility must be taken into account, the non-geometric approach reported in Ref.~\onlinecite{diaz} can also be used. In fact, we must just follow the steps described therein, but we must be careful if there are reducibility constraints. Basically, what we have to do is to check if for some election of the gauge parameters, the gauge transformations of the field variables vanish on-shell (they become then trivial gauge transformations), which means that the parameters are not independent among themselves. This fact has to be taken into account in the counting of the gauge parameters $e$. If this were the case, we must also count independent constraints only, of course. 

(g) It might happen that for some theories we have some information relevant for the counting from the very beginning. For example, sometimes we already know from outset the gauge symmetries of a given theory.  In that case, in order to use the formula \eqref{gl1}, we are only missing the number of Lagrangian constraints, which can be found by the algorithm reported here, or the non-geometric one reported in Ref.~\onlinecite{diaz}. Of course, in that case, we have to be sure that we have all the independent gauge transformations. 

Finally, further work is needed to see how the Lagrangian approach works with fermions (in both point particle systems and field theory) and to compare it with the case of bosonic variables developed in Ref.~\onlinecite{diaz} and in this paper. 

\begin{acknowledgments}
We thank Mariano Celada, Gerardo F. Torres del Castillo, Mercedes Vel\'azquez, and Jos\'e D. Vergara for their valuable comments on this work. Merced Montesinos' sabbatical year at {\it Benem\'erita Universidad Aut\'onoma de Puebla} was supported by CONACYT, Grant No. 266062 and by {\it Benem\'erita Universidad Aut\'onoma de Puebla}. 
\end{acknowledgments}

%%%%%%%%%%%%%%%%%%%%%%%%%%%%%%%%%%%%%%%%%%%%
\appendix
\section{TOY MODEL}\label{apen1}
In this appendix, we present a simple toy model, which allows us to illustrate some technicalities of the general theory that are not present in the physical examples analyzed in Sect. \ref{sec3} of this paper.  In particular, it is displayed why in some cases linear combinations of the constraints in Eq. (\ref{vecf}) must be considered. The model is defined by the action 
\begin{eqnarray}
S[\phi^a]&=&\int_M \d^n x \left[ \frac{\left(\dot{\phi}^2\right)^2}{2}+ \frac{\left(\dot{\phi}^3\right)^2}{2} + \phi^4\left( \phi^1-\phi^2 \right)+\phi^5\left( \phi^1-\phi^3 \right) \right], \label{ejem}
\end{eqnarray}
where $a=1,2,3,4,5$. Notice that we could have considered the corresponding point mechanical system, but we want to keep for consistency the field theory formalism along all the work (we remind the reader that the approach reported in this paper works perfectly for both field theory and point particle systems). The variation of (\ref{ejem}) with respect to the configuration variables yields the equations of motion
\begin{eqnarray}
\delta \phi^1 &:&  \phi^4+\phi^5 =0 ,\nonumber\\
\delta \phi^2 &:&  \ddot{\phi}^2+\phi^4 =0 ,\nonumber\\
\delta \phi^3 &:&  \ddot{\phi}^3+\phi^5=0 ,\nonumber\\
\delta \phi^4 &:&  \phi^1-\phi^2=0 ,\nonumber\\
\delta \phi^5 &:&  \phi^1-\phi^3=0 .\label{eqapen}
\end{eqnarray}

{\it Lagrangian constraints}. The points of the $10$-dimensional velocity phase space are labeled by ($\phi^a, {\dot \phi}^a$). From \eqref{ejem}, it follows that the corresponding symplectic structure (\ref{gp1}), energy (\ref{energy}), and the differential of the energy are
\begin{eqnarray}\label{seapen}
\Omega &=&\int \d^{n-1}x \left( \ed \dot{\phi}^2 \ew \ed \phi^2+ \ed \dot{\phi}^3 \ew \ed \phi^3 \right), \nonumber\\
E &=&\int \d^{n-1}x  \left[ \frac{\left(\dot{\phi}^2\right)^2}{2}+ \frac{\left(\dot{\phi}^3\right)^2}{2}- \phi^4\left( \phi^1-\phi^2 \right) -\phi^5\left( \phi^1-\phi^3 \right) \right], \nonumber \\
\ed E&=& \int \d^{n-1}x \left[ \dot{\phi}^2 \ed \dot{\phi}^2+ \dot{\phi}^3 \ed \dot{\phi}^3+ \phi^4 \ed \phi^2+ \phi^5\ed \phi^3 - \left( \phi^4+\phi^5 \right) \ed \phi^1  \right. \nonumber\\
&&- \left( \phi^1-\phi^2 \right) \ed \phi^4 \left. - \left( \phi^1-\phi^3 \right) \ed \phi^5\right].
\end{eqnarray} 
Therefore, if we write the vector field $X$ as
\begin{equation}
X:=\int \d^{n-1}x \left( \alpha^a\frac{\delta}{\delta \phi^a }+\beta^a \frac{\delta}{\delta\dot{\phi}^a}\right),
\end{equation}
then Eq. (\ref{gp2+}) becomes
\begin{align}
&&\alpha^2 = \dot{\phi}^2, && 0= \phi^4+\phi^5,\nonumber\\
&&\alpha^3 = \dot{\phi}^3, && 0= \phi^1-\phi^2, \nonumber\\
&&\beta^2  = -\phi^4, && 0= \phi^1-\phi^3, \nonumber \\
&&\beta^3  = -\phi^5, && \label{Leapen}
\end{align}
which of course coincide with (\ref{eqapen}), once we substitute $\alpha^a= \dot{\phi}^a$ and $\beta^a= \ddot{\phi}^a$ therein. 

On the other hand, the rank of the symplectic structure (\ref{seapen}) is four. In fact, a basis of $\ker \Omega$ is given by 
\begin{eqnarray}
Z^1_{i}&:=& \int \d^{n-1}x \, \frac{\delta}{\delta \phi^i}, \qquad Z^2_{i}:= \int \d^{n-1}x\, \frac{\delta}{\delta \dot{\phi}^i}, \label{nullape}
\end{eqnarray}
where $i=1,4,5$. Nevertheless, only $Z^1_{i}$ generates the Lagrangian constraints given by
\begin{eqnarray}
\int \d^{n-1}x  \,\varphi_{1} & :=& Z^1_{1}\cdot \ed E =\int \d^{n-1}x \left(\phi^4+\phi^5\right)\approx 0, \nonumber\\
\int \d^{n-1}x \, \varphi_{2} & :=& Z^1_{4}\cdot \ed E= \int \d^{n-1}x \left( \phi^1-\phi^2 \right) \approx 0,\nonumber\\
\int \d^{n-1}x \, \varphi_{3} & :=& Z^1_{5}\cdot \ed E= \int \d^{n-1}x \left( \phi^1-\phi^3 \right) \approx 0.
\end{eqnarray}
%Notice that these constraints are part of the equations of motion (\ref{eqapen}). 
Continuing with the approach, we must demand that $X\left( \int \d^{n-1}x\varphi_{i} \right) \approx 0$. Since $X$ satisfies \eqref{Leapen}, we have
\begin{eqnarray}
0&\approx& X\left(\int \d^{n-1}x \, \varphi_1\right)=\int \d^{n-1}x\left(\alpha^4 + \alpha^5\right),\nonumber\\
0&\approx& X\left(\int \d^{n-1}x \, \varphi_2\right)=\int \d^{n-1}x\left(\alpha^1 - \dot{\phi}^2 \right) ,\nonumber\\
0&\approx& X\left(\int \d^{n-1}x \, \varphi_3\right)=\int \d^{n-1}x\left(\alpha^1 - \dot{\phi}^3\right). \label{newape}
\end{eqnarray}
Notice that these expressions involve components of $X$ that are not fixed by (\ref{Leapen}). These are the relations predicted by the general theory of Sec. \ref{sec2}, and at this stage, they are not Lagrangian constraints because they are not functions of $\left(\phi^a, \dot{\phi}^a\right)$ (they depend on $\alpha^1$, $\alpha^4$, and $\alpha^5$ too). So, we have to handle relations (\ref{newape}) in order to get rid of the $\alpha$'s and get more Lagrangian constraints. 

We find that if we use the second and the third equations of (\ref{newape}) we can eliminate the variable $\alpha^1$, obtaining a relation between the field variables only. In fact, we have
\begin{eqnarray}
0&\approx& \int \d^{n-1}x \left[ \alpha^1 - \dot{\phi}^2-\left( \alpha^1 - \dot{\phi}^3 \right) \right] \nonumber\\
&=&\int \d^{n-1}x\left( \dot{\phi}^3 -\dot{\phi}^2\right)=: \int \d^{n-1}x \,  \varphi_4, 
\end{eqnarray} 
which is a new Lagrangian constraint. Now, we also must demand that $X$ overrides this constraint, i.e., 
\begin{eqnarray}
0\approx X\left( \int \d^{n-1}x \, \varphi_4 \right)= \int \d^{n-1}x \left( \phi^4-\phi^5 \right)=: \int \d^{n-1}x \, \varphi_5, 
\end{eqnarray}
which implies the new Lagrangian constraint $\varphi_5$. Finally, we demand $X$ to be tangent to this constraint, and we obtain 
\begin{equation}
0\approx X\left( \int \d^{n-1}x  \, \varphi_5 \right)= \int \d^{n-1}x \left( \alpha^4-\alpha^5 \right) . \label{ucap}
\end{equation}
Notice that this relation also contains components of the vector field $X$ that were not fixed by the equations of motion. This is the end of the constraint algorithm. Therefore, we have $5$ projectable Lagrangian constraints $\varphi_1, \varphi_2, \varphi_3, \varphi_4,$ and $\varphi_5$ and  the relations \eqref{newape} and \eqref{ucap}.

To continue our analysis, we must use the requirement $\alpha^a=\dot{\phi}^a$ (that arises by the second-order equation problem) in \eqref{Leapen} and in the relations \eqref{newape} and \eqref{ucap}. Notice that from \eqref{Leapen} we do not obtain any constraint, and using it in \eqref{newape} and \eqref{ucap} gives
\begin{eqnarray}\label{otramas}
\varphi_6&:=& \int \d^{n-1}x\left(\dot{\phi}^4 + \dot{\phi}^5\right),\qquad \varphi_7:= \int \d^{n-1}x\left(\dot{\phi}^1 - \dot{\phi}^2 \right),\nonumber\\
\varphi_8&:=& \int \d^{n-1}x\left(\dot{\phi}^1 - \dot{\phi}^3 \right),\qquad \varphi_9:= \int \d^{n-1}x \left(\dot{\phi}^4-\dot{\phi}^5 \right).
\end{eqnarray}
respectively. These are non-projectable constraints. Notice that because of $\varphi_4$ we have that $\varphi_7$ and $\varphi_8$ are not new independent constraints. Indeed, $\varphi_7-\varphi_8=\varphi_4$. Therefore, we have only $3$ independent non-projectable constraints. Therefore, the toy model has $l=5+3=8$ independent Lagrangian constraints.

\subsection{Degree of freedom count}

{\it Associated vector fields}. It is clear, as predicted by the general theory, that the non-projectable constraints (\ref{otramas})  do not have associated vector fields because of the variables $\dot{\phi}^4,\dot{\phi}^5$ and $\dot{\phi}^1$ therein contained. Note that these variables do not appear in the symplectic structure (\ref{seapen}) and they cannot be eliminated by combining the constraints (\ref{otramas}). For the projectable constraints we define the following smeared constraints
\begin{eqnarray}
\varphi_1[N_1]&=& \int \d^{n-1}x \, N_1 \left(\phi^4+\phi^5 \right),\qquad \varphi_2[N_2]= \int \d^{n-1}x \, N_2 \left(\phi^1-\phi^2 \right) ,\nonumber\\
\varphi_3[N_3]&=&\int \d^{n-1}x  \, N_3 \left(\phi^1-\phi^3 \right),\qquad \varphi_4[N_4]=\int \d^{n-1}x \, N_4 \left(\dot{\phi}^3 -\dot{\phi}^2 \right),\nonumber\\
\varphi_5[N_5]&=&\int \d^{n-1}x \,  N_5 \left(\phi^4 - \phi^5 \right).
\end{eqnarray}
Now, we must establish Eq. (\ref{vecf}). First, we establish it for each projectable constraint. However, for the constraints $\varphi_1, \varphi_2, \varphi_3$, and $\varphi_5$, there does not exist any vector field that satisfies Eq. (\ref{vecf}) because of the presence of $\phi^1, \phi^4$, and $\phi^5$ therein. For the constraint $\varphi_4$, the associated vector field is
\begin{equation}
Z_4[N_4]= \int \d^{n-1} x \, N_4\left(\frac{\delta}{\delta \phi^2}-\frac{\delta}{\delta \phi^3}\right).
\end{equation}
Now, we establish Eq. (\ref{vecf}) by taking two constraints. A key observation is that although the constraints $\varphi_2$ and $\varphi_3$ do not possess associated vector fields because of the presence of $\phi^1$, this variable can be eliminated by combining them. We have that 
\begin{eqnarray}
\varphi_{10}[M]&:=&\varphi_2[M]+\varphi_3[-M]\nonumber\\
&=& \int \d^{n-1}x\left[  M \left(\phi^1-\phi^2   \right)+ \left(-M\right) \left(\phi^1-\phi^3\right) \right]\nonumber\\
&=& \int \d^{n-1} x \, M \left( \phi^3-\phi^2\right) ,
\end{eqnarray}
where we have taken $M$ and $-M$ as smearing functions in the constraints. Now, we observe that all variables that appear in $\varphi_{10}[M]$ are present in the symplectic structure \eqref{seapen}, and it can readily be seen that its associated vector field is
\begin{equation}
Z_{10}[M]= \int \d^{n-1} x  \, M \left( -\frac{\delta}{\delta \dot{\phi}^2}+\frac{\delta}{\delta \dot{\phi}^3}\right).
\end{equation}
Now, we can try to establish Eq. (\ref{vecf}) by taking other combinations that involve other two, three, or the four constraints. However, no other combination gives an independent associated vector field. Therefore, Eq. (\ref{vecf}) gives two associated vector fields. Notice, that if we had considered each constraint separately we would have concluded that Eq. (\ref{vecf}) only could be established once. This information is vital in the counting of the physical degrees of freedom. Notice that this is the procedure that we can follow in practice and we must verify if the variables that prevent the constraints from having associated vector fields can be eliminated by combining them.

{\it Null associated vector fields}. Now, using the associated vector fields $Z_4, Z_{10}$ and the null vectors (\ref{nullape}), we must check if we can form vector fields that are tangent to the Lagrangian constraint surface. Therefore, we form the vector fields
\begin{eqnarray}
Z&=& Z_4[N_4]+Z^1_{i}[P^i]+ Z^2_{i}[Q^i], \nonumber\\
Z'&=& Z_{10}[M]+Z^1_{i}[R^i]+ Z^2_{i}[S^i].
\end{eqnarray}
The requirement of being tangent to the Lagrangian constraint surface yields 
\begin{align}
&P^{1}\approx 0, &&N_4 \approx 0, && P^4 \approx 0, \nonumber \\
&P^{5}\approx 0, &&Q^{1} \approx 0, &&Q^{5} \approx -Q^4,\nonumber\\
&R^1\approx 0, &&  M \approx 0, && R^4 \approx 0, \nonumber \\
&R^{5} \approx 0 , &&S^{1} \approx 0, &&S^{5} \approx -S^4,
\end{align}
where $Q^4$ and $S^4$ are left arbitrary. We only obtain a relation between parameters that smear the original null vectors.
As a consequence, the unique tangent null vector is a combination of two of the original null vector, for example, $Z^1_{4}[Q^4]+Z^1_{5}[-Q^4]$. Therefore, the vectors $Z_4$ and $Z_{10}$ do not generate null associated vector fields. This is in agreement with the general theory that says that only the vectors that are associated with constraints that are part of the equations of motion generate any null associated vectors fields. Then, we do not have any null associated vector field. 

{\it Degree of freedom count}. We are ready to make the counting of the physical degrees of freedom: the number of field variables $\phi^a$ is $N=5$, the total number of null vectors, given by \eqref{nullape}, is $B = 6$, and the number of linearly independent combinations of constraints to get the associated vector fields is $C =2$. Therefore, using \eqref{pdfc}, the number of physical
degrees of freedom is
\begin{equation}
\frac{1}{2} {\rm Rank} \, \iota^*\Omega = 5-\frac{1}{2} \left(6+2 \right) =1. \label{glapen}
\end{equation}
Then, this model has locally $1$ degree of freedom.  

Furthermore, by substituting this result into the right-hand side of Eq. \eqref{gl1}, we get $g + e = 2(N-1)-l= 2(5-1)-8=0$. Therefore, $\iota^*\Omega$ does not have null vector fields.
Notice that because $g+e=0$, then $g=0=e$. This means that the model has no gauge freedom. In Appendix \ref{A2}, we will see how the current Lagrangian approach to find the gauge symmetries is in agreement with this result.

On the other hand, notice that in this example it is actually possible to calculate $\iota^*\Omega$. In fact, we have $\iota^*\Omega\approx 2\int \d^{n-1}x \, \, \ed \dot{\phi}^2 \ew \ed \phi^2$, from which we directly have $\frac{1}{2} {\rm Rank} \, \iota^*\Omega=1$. Therefore, in this case we could have calculated the physical degrees of freedom without looking for the associated vector fields, etc. As we explained in the general theory, this fact (that is always possible for point particles) is only possible for some particular field theories. 

\subsection{Off-shell gauge transformations}\label{A2}

Following the steps explained in Sec. \ref{GT}, first we use only the null vector field that is tangent to the Lagrangian constraint surface $Z^1_{4}[\varepsilon]+Z^1_{5}[-\varepsilon]$, where $\varepsilon$ is a smearing function. We form the vector field
\begin{eqnarray}
\tilde{X}&:=&\int \d^{n-1} x \, \varepsilon \left(\frac{\delta}{\delta \dot{\phi}^4}-\frac{\delta}{\delta \dot{\phi}^5}\right). \label{gauapen}
\end{eqnarray}
Second, we demand the vector to be casted in the form
\begin{eqnarray}
\tilde{X}&=&\int \d^{n-1}x \left[ \delta_{\varepsilon}\phi^a \frac{\delta}{\delta\phi^a}+\partial_0(\delta_{\varepsilon} \phi^a) \frac{\delta}{\delta \dot{\phi}^a} \right],
\end{eqnarray}
which yields $\varepsilon= 0$. Consequently, the gauge vector field reduces to zero. Therefore, the model does not have any gauge symmetry. This is in agreement with the already mentioned fact that $e=0=g$.

Finally, the map (\ref{map}) allows us to know the number of first- and second-class constraints $N_1 = e = 0$, and $N_2 = l + g - e = 8+0-0 =8$, which should appear if Dirac's canonical analysis of the toy model is performed.

\section{RELATION WITH THE COVARIANT CANONICAL FORMALISM}\label{apen2}

In this appendix, we show the relationship between the symplectic structure reported in this paper (\ref{gp1}) and that of the CCF.  The main difference is conceptual. While our symplectic structure (\ref{gp1}) is defined on $T \mathcal{C}$, that of the CCF is defined on the space of field configurations ${\mathcal F}$ and it is given by \cite{crnkovic,bombelli,Wald}
\begin{eqnarray}\label{lee+wald}
\omega &=& \int_\Sigma \d\sigma_{\mu} \left[   \frac{\partial^2 {\mathcal L}}{\partial \phi^A \partial \phi^B_{\mu}} \,  \ed \phi^A  \ew  \ed \phi^B   + \frac{\partial^2 {\mathcal L}}{\partial {\phi}^A_{\nu} \partial \phi^B_{\mu}} \left(\partial_{\nu}\, \ed\phi^A\right) \ew \ed\phi^B \right ],
\end{eqnarray}
where $\Sigma$ is a space-like surface, $\d \sigma_{\mu}= n_{\mu} \d\sigma$ with $\d \sigma$ being the volume element of $\Sigma$ and $n_{\mu}$ being the unit normal. Note that its action on two vectors in the tangent space to ${\mathcal F}$, $ X_1 = \int_{\Sigma} \d \sigma \, \delta_1 \phi^A  \frac{\delta}{\delta \phi^A}$ and $X= \int_{\Sigma} \d \sigma\, \delta_2 \phi^B \frac{\delta}{\delta \phi^B}$, is
\begin{eqnarray}
\omega ( X_1, X_2)= \int_\Sigma  \d\sigma_{\mu} \, \omega^\mu, \label{ssccf}
\end{eqnarray}
with
\begin{eqnarray}
\omega^{\mu}&=& \frac{\partial^2 {\mathcal L}}{\partial \phi^A \partial \phi^B_{\mu}} \left[ \delta_1\phi^A \delta_2\phi^B - \delta_1\phi^B \delta_2\phi^A\right]  \nonumber\\
&& + \frac{\partial^2 {\mathcal L}}{\partial {\phi}^A_{\nu} \partial \phi^B_{\mu}} 
%\nonumber \\&& \times
 \left[ \left(\partial_{\nu} \delta_1\phi^A\right) \delta_2\phi^B - \delta_1\phi^B \left(\partial_{\nu}\delta_2\phi^A\right) \right].
 \end{eqnarray}
To make this calculation, we have used that $\ed \phi^B( x'^{\mu}) \left( \frac{\delta}{\delta \phi^A (x^{\mu})} \right)=\frac{\delta}{\delta \phi^A(x^{\mu})} \left( \phi^B( x'^{\mu}) \right) := \frac{\delta \phi^B( x'^{\mu})}{\delta \phi^A (x^{\mu})} := \delta^{B}_A \delta( x'^{\mu}-x^{\mu})$, \cite{crnkovic} where we have restored the spacetime dependence. The definition of $\omega$ depends on $\Sigma$ generically. Only in the case when the vector fields $X_1$ and $X_2$ are solutions to the linearized equations of motion, $\omega$ is independent of $\Sigma$ provided that either $\Sigma$ is compact or the fields satisfy suitable asymptotic conditions to ensure that no spatial boundary terms arise from applying Gauss theorem. \cite{Wald, woodhouse} It is usual in the CCF to restrict the analysis to the space of solutions $\bar{\mathcal{F}}\subset \mathcal{F}$ and act their symplectic structure only on vector fields that are solution to the linearized equations of motion.

By comparing \eqref{lee+wald} and \eqref{gp1}, we conclude that \eqref{lee+wald}, with a different interpretation, reduces to \eqref{gp1} if we choose $\Sigma$ in (\ref{lee+wald}) to be a $\Sigma_t$ of the foliation of spacetime because with this choice we have $n_{\mu}=(1,0,\ldots, 0)$ and so $ \d \sigma = \d^{n-1} x$. Due to the fact that $\omega$ and $\Omega$ are defined on two different spaces, the tangent vectors to these spaces are also different from each other. For instance, the symplectic structure \eqref{gp1} can act over arbitrary vector fields $X= \int \d^{n-1}x \left(\alpha^A \frac{\delta}{\delta \phi^A} + \beta^A \frac{\delta}{\delta \dot \phi^A} \right)$ that are tangent to the velocity phase space. \cite{woodhouse}

Let us consider the three dimensional generalized Palatini action \eqref{palati} to illustrate the differences between the symplectic structure considered in this paper (\ref{gp1}) and that of the CCF. For this theory we have that the symplectic structure is given by \eqref{psp} 
\begin{eqnarray}
\Omega &=&-\int \d^{2}x \, \epsilon^{0ab} \ed e_{ia} \ew \ed A^i_b,
\end{eqnarray}
which is defined on a velocity phase space of dimension $12 \times{\rm dim} \, \mathfrak{g}$ and has $8 \times{\rm dim} \, \mathfrak{g}$ null vectors given by (\ref{inv}), as we already mentioned. 

On the other hand, the corresponding symplectic structure of the CCF, given by (\ref{lee+wald}), acquires the form
\begin{eqnarray}\label{mensos}
\omega &=&- \int_{\Sigma}  \d\sigma_{\mu} \, \epsilon^{\mu\nu\rho} \ed e_{i\nu} \ew \ed A^i_{\rho}, \label{SSCCF}
\end{eqnarray}
which is defined on a $(6 \times{\rm dim} \, \mathfrak{g})$-dimensional field configuration space ${\mathcal F}$. The degeneracy directions of $\Omega$ consist of those field variations for which $\delta e_{i\nu}$ and $\delta A^i_{\rho}$ vanish on $\Sigma$.

As a side comment, if we restrict ourselves as is usually done to the space of solutions $\bar{\mathcal{F}}$, the symplectic structure \eqref{SSCCF} possesses the following null vector fields:
\begin{eqnarray}
\tilde{X}&=&\int_{\Sigma} \d \sigma \left[ \left( \mathcal{D}_{\mu} \rho^i+f^i{}_{jk} e^j_{\mu} \tau^k \right)\frac{\delta}{\delta e^i_{\mu}}+ \left(\mathcal{D}_{\mu}\tau^i - \Lambda \epsilon^i{}_{jk} e^j_{\mu}\rho^k \right) \frac{\delta}{\delta A^i_{\mu}} \right].
\end{eqnarray}

%//////////////////////////////////////////////////////////////
\nocite{*}
%\bibliography{aipsamp}% Produces the bibliography via BibTeX.

\begin{thebibliography}{100}

\bibitem{Dirac1} P. A. M. Dirac,  ``Generalized Hamiltonian dynamics," \href{http://dx.doi.org/10.4153/CJM-1950-012-1}{Can. J. Math.} {\bf 2},  129 (1950);  ``Generalized Hamiltonian dynamics," \href{http://rspa.royalsocietypublishing.org/content/246/1246/326.short}{Proc. R. Soc. London, Ser. A} {\bf 246},  326 (1958).

\bibitem{Dirac2} P. A. M. Dirac,  \emph{Lectures on Quantum Mechanics} (Belfer Graduate School of Science, New York, 1964).

\bibitem{crnkovic} C. Crnkovic and E. Witten, ``Covariant description of canonical formalism in geometrical theories," in {\it Three Hundred Years of Gravitation}, edited by S. Hawking and W. Israel (Cambridge University Press, Cambridge, 1987). 

\bibitem{bombelli} A. Ashtekar, L. Bombelli, and O. Reula, ``The covariant phase space of asymptotically flat gravitational fields," in {\it Mechanics, Analysis, and Geometry: 200 years after Lagrange}, edited by M.
Francaviglia and D. Holm (North-Holland, Amsterdam, 1990).
 
\bibitem{Wald} J. Lee and R. M. Wald, ``Local symmetries and constraints," \href{http://dx.doi.org/10.1063/1.528801}{J. Math. Phys.} {\bf 31}, 725 (1990).
 
\bibitem{Gotay} M. J. Gotay and J. M. Nester,  ``Presymplectic Lagrangian systems I: the constrain algorithm and the equivalence theorem," Ann. Inst. Henri Poincar\'e , Sect. {\bf A 30}, 129 (1979), available at  \href{https://eudml.org/doc/76022}{http://eudml.org/doc/76022}.

\bibitem{Batlle} C. Batlle, J. Gomis, J. M. Pons, and N. Roman, ``Equivalence between the Lagrangian and Hamiltonian formalism for
constrained systems," \href{http://dx.doi.org/10.1063/1.527274}{J. Math. Phys.} {\bf 27}, 2953 (1986).

\bibitem{GotayNester} M. J. Gotay and J. M. Nester,  ``Presymplectic Lagrangian systems II: the second-order equation problem," Ann. Inst. Henri Poincar\'e, Sect. {\bf A 32},  1 (1980), available at \href{https://eudml.org/doc/76059}{https://eudml.org/doc/76059}.
 
\bibitem{Hinds} M. J. Gotay, J. M. Nester, and G. Hinds,  ``Presymplectic manifolds and the Dirac-Bergmann theory of constraints," \href{http://dx.doi.org/10.1063/1.523597}{J. Math. Phys.} {\bf 19}, 2388 (1978).

\bibitem{diaz} B. D\'iaz, D. Higuita, and M. Montesinos,  ``Lagrangian approach to the physical degree of freedom count," \href{http://dx.doi.org/10.1063/1.4903183}{J. Math. Phys.} {\bf 55}, 122901 (2014).

\bibitem{Klein}  R. Klein, and D. J. Roest,  ``Exorcising the Ostrogradsky ghost in coupled systems," \href{http://dx.doi.org/10.1007/JHEP07(2016)130}{JHEP} {\bf 07}, 130 (2016).

\bibitem{Klein2} M. Crisostomi, R. Klein, and D. Roest, ``Higher derivative field theories: degeneracy conditions and classes," \href{https://doi.org/10.1007/JHEP06(2017)124}{JHEP} {\bf 06}, 124 (2017).

\bibitem{SunderSy} K. Sundermeyer, {\it Symmetries in Fundamental Physics},   Fundam. Theor. Phys. 176 (Springer International Publishing,, 2014).

\bibitem{woodhouse} N. M. J. Woodhouse, {\it Geometric Quantization}, 2nd ed. (Oxford University Press, Oxford, 1991).
 
\bibitem{Nester}J. M. Nester,  ``Invariant derivation of the Euler-Lagrange equation," \href{http://iopscience.iop.org/0305-4470/21/21/003}{J. Phys. A: Math. Gen.} {\bf 21}, L1013 (1988).

\bibitem{Pons} J. M. Pons,  ``New relation between Hamiltonian and Lagrangian constraints," \href{http://iopscience.iop.org/0305-4470/21/12/014/}{J. Phys. A: Math. Gen.} {\bf 21}, 2705 (1988).

\bibitem{Sundermeyer} K. Sundermeyer, {\it Constrained Dynamics}. Lecture Notes in Physics Vol. {\bf 169} (Springer-Verlag, Berlin, 1982).

\bibitem{Teitelboim} M. Henneaux and C. Teitelboim, {\it Quantization of Gauge Systems} (Princeton University Press, Princeton, 1992).

\bibitem{note1} For example, consider the symplectic structure $\Omega=\int \d^{n-1}x \left(\ed\phi^1\ew \ed\dot{\phi}^1+ \ed\phi^2\ew \ed\dot{\phi}^2 \right)$ defined on a velocity phase space whose points are locally labeled by $\{\phi^1,\phi^2,\dot{\phi}^1,\dot{\phi}^2\}$, and the constraints $\varphi_1[M]=\int \d^{n-1}x \, M \phi^2 \approx 0$ and $\varphi_2[N]=\int \d^{n-1}x\, N \dot{\phi}^2 \approx 0$, where $M$ and $N$ are smearing functions. Notice that the vector fields $X_1= -\int \d^{n-1}x \, M \frac{\delta}{\delta \dot{\phi}^2}$ and $X_2= \int \d^{n-1}x \, N \frac{\delta}{\delta \phi^2}$ satisfy $X_1\cdot \Omega= \ed\left(\varphi^2[M] \right)$ and $X_2\cdot \Omega= \ed \left(\varphi^2[N] \right)$, respectively. However, as is clear, these vectors are not tangent to the constraint surface $X_1(\varphi_2[N])=\int \d^{n-1}x \,MN =X_2\left(\varphi_1[M]\right)$. So, there are no null associated vector fields. Nevertheless, these constraints change the rank of $\Omega$. Therefore, we must take this into account when we calculate it. In fact, we have ${\rm rank} \,\iota^*\Omega= 4-2=2$, where $4$ is the original rank and we subtract $2$ because we have two associated vector fields [i.e., two relations of the type \eqref{vecf}]. Notice that in this case we have $\iota^*\Omega\approx  \int \d^{n-1}x \left( \ed\phi^1\ew \ed\dot{\phi}^1 \right)$.

\bibitem{note2} For example, consider the symplectic structure $\Omega=\int \d^{n-1}x \left( \ed\phi^1\ew \ed \dot{\phi}^1\right)$ defined on a velocity phase space whose points are locally labeled by $\{\phi^1, \phi^2,\dot{\phi}^1,\dot{\phi}^2\}$. Now, consider the constraint $\varphi_1[M]= \int \d^{n-1}x\, M \phi^2 \approx 0$. This constraint reduces (by the inclusion) the dimension of the space in which $\Omega$ is defined by one, except it does not change its rank. In contrast, consider the constraint $\varphi_2[N]=\int \d^{n-1}x\, N \phi^1 \approx 0$, which turns $X_2=-\int \d^{n-1}x\, N\frac{\delta}{\delta \dot{\phi}^1}$ into an associated vector, because $ X_2\cdot \Omega= \ed \left(\varphi_2[N]\right)$. Also, it overrides the constraints $\varphi_1[M]$ and $\varphi_2[N]$. Then, the vector $X_2$ is a null associated vector. So, the counting of rank $\iota^*\Omega$ takes into account the original two null vector fields $\int \d^{n-1}x \, \frac{\delta}{\delta \phi^2}, \int \d^{n-1}x\, \frac{\delta}{\delta \dot{\phi}^2}$, the constraint $\varphi_2$, and $X_2$. Therefore, the rank of $\Omega$ after the inclusion is $4-(1+1+1+1)=0$ (which can also be verified directly, $\iota^*\Omega\approx 0$).

\bibitem{Noether} E. Noether,  ``Invariante variations probleme," Nachr. d. K\"{o}nig. Gesellsch. d. Wiss. zu G\"{o}ttingen, Math-phys. Klasse {\bf 1918},  235, available at \href{https://eudml.org/doc/59024}{https://eudml.org/doc/59024}.

\bibitem{translation} E. Noether, ``Invariant variation problems," \href{http://dx.doi.org/10.1080/00411457108231446}{Transp. Theory and Stat. Phys.} {\bf 1},  186 (1971).

\bibitem{Beng} I. Bengtsson, ``Hamiltonian treatment of free string field theory," \href{https://doi.org/10.1016/0370-2693(86)90265-0}{Phys. Lett. B} {\bf 172}  342 (1986).

\bibitem{Gomis} C. Batlle, and J. Gomis, ``Hamiltonian formalism of the gauge invariant free closed bosonic string field theory," \href{https://doi.org/10.1016/0370-2693(87)90072-4}{Phys. Lett. B} {\bf 187} 61 (1987).

\bibitem{Blau} M. Blau and G. Thompson, ``Topological gauge theories of antisymmetric tensor fields," \href{https://doi.org/10.1016/0003-4916(91)90240-9}{Ann. Phys.} {\bf 205}  130 (1991).

\bibitem{Thompson}  D. Birmingham, M. Blau, M. Rakowski and G. Thompson, ``Topological field theory," \href{https://doi.org/10.1016/0370-1573(91)90117-5}{Phys. Rep.} {\bf 209}  129 (1991).

\bibitem{Caicedo} M. I. Caicedo and A. Restucia,  ``On the BRST charge for topological BF theories," \href{http://www.sciencedirect.com/science/article/pii/037026939390195N}{Phys. Lett. B} {\bf 307}, 77 (1993).

\bibitem{Cuesta} V. Cuesta, M. Montesinos, M. Vel\'azquez, and J. D. Vergara, ``Topological field theories in $n$-dimensional spacetimes and Cartan's equations," \href{https://journals.aps.org/prd/abstract/10.1103/PhysRevD.78.064046}{Phys. Rev. D} {\bf 78}, 064046 (2008).

\bibitem{Celada} M. Montesinos, D. Gonz\'alez, M. Celada, and B. D\'{\i}az,  ``Reformulation of the symmetries of first-order general relativity," \href{https://doi.org/10.1088/1361-6382/aa89f3}{Classical Quantum Gravity} {\bf 34}, 205002 (2017).

\bibitem{Romano} J. D. Romano,  ``Geometrodynamics vs. connection dynamics," \href{http://dx.doi.org/10.1007/BF00758384}{Gen. Relativ. Gravit.} {\bf 25},  759 (1993).

\bibitem{Review} M. Celada, D. Gonz\'alez, and M. Montesinos, ``$BF$ gravity," \href{https://doi.org/10.1088/0264-9381/33/21/213001}{Classical Quantum Gravity} {\bf 33},   213001 (2016).

\bibitem{Witten} E. Witten,  ``2 + 1-dimensional gravity as an exactly soluble system," \href{http://dx.doi.org/10.1016/0550-3213(88)90143-5}{Nucl. Phys.} {\bf B311}, 46 (1988).

\bibitem{Livine} V. Bonzom and E. Livine,   ``A Immirzi-like parameter for 3d quantum gravity," \href{http://dx.doi.org/10.1088/0264-9381/25/19/195024}{Classical Quantum Gravity} {\bf 25},  195024 (2008).

\bibitem{Chern} S.-S. Chern and J. Simons,  ``Characteristic forms and geometric invariants," \href{http://doi.org/10.2307/1971013}{Annals Math.} {\bf 99},  48 (1974).

\bibitem{Naka} M. Nakahara, {\it Geometry, Topology and Physics} (Institute of Physics Publishing, Bristol 
and Philadelphia, 1990).

\bibitem{Bla} M. Blagojevi\'{c}, {\it Gravitation and Gauge Symmetries} (Institute of Physics
Publishing, Bristol and Philadelphia, 2002).

\bibitem{Mondragon} M. Mondrag\'on and M. Montesinos,  ``Covariant canonical formalism for four-dimensional BF theory," \href{http://dx.doi.org/10.1063/1.2161805}{J. Math. Phys.} {\bf 47}, 022301 (2006).

\bibitem{Mond} M. Mondrag\'on, ``Comment on dimension of the moduli space and Hamiltonian analysis of BF field theories," \href{http://www.worldscientific.com/doi/abs/10.1142/S0217751X12750012}{Int. J. Mod. Phys. A} {\bf 27}, 1275001 (2012).

\bibitem{Contreras} M. Ba\~nados and M. Contreras,  ``Darboux coordinates for (first-order) tetrad gravity," \href{http://dx.doi.org/10.1088/0264-9381/15/6/009}{Classical Quantum Gravity} {\bf 15},  1527 (1998).

\bibitem{Contreras2} M. Contreras and J. Zanelli, ``A note on the spin connection representation of gravity," \href{https://doi.org/10.1088/0264-9381/16/6/334}{Classical Quantum Gravity} {\bf 16}, 2125 (1999).

\end{thebibliography}

\end{document}